\documentclass[11pt]{article}

\usepackage[papersize={17cm,24cm},margin=22mm,top=17mm,headsep=5mm]{geometry} 

\usepackage{verbatim} 
\usepackage{titling} 
\usepackage{csquotes} 
\usepackage{upquote} 
\usepackage[bottom]{footmisc} 
\usepackage{enumitem} 
\setitemize{itemsep=2pt,topsep=4pt,parsep=0pt,partopsep=0pt}
\usepackage{covington}  

\usepackage[english]{babel} 

\date{}
\usepackage[breaklinks,colorlinks,urlcolor=black,citecolor=black,linkcolor=black]{hyperref} 

\usepackage[compact,noindentafter]{titlesec} 
\titlespacing{\section}{0pt}{2.7mm}{1.5mm} 
\titlespacing{\subsection}{0pt}{2mm}{1pt}
\titlespacing{\subsubsection}{0pt}{1.2mm}{0.5pt}

\newcommand\blfootnote[1]{
  \begingroup
  \renewcommand\thefootnote{}\footnote{\hspace{-2.4em} #1}%
  \addtocounter{footnote}{-1}
  \endgroup
}

\renewcommand{\maketitle}
  {\bgroup\setlength{\parindent}{0pt}
   \begin{flushleft}
    \textbf{\LARGE{\ \\ \vspace{20mm}\strut\thetitle\strut}\vspace{2mm}}
    
    {\Large\theauthor}
    \vspace{2mm}\\
    \textit{Universit\'e Paris-Saclay, CNRS,  Laboratoire Aim\'e Cotton, 91405, Orsay, France}

    \vspace{2mm}
    \text{19.07.2024}
  \end{flushleft}\egroup\vspace{18mm}
}

\renewenvironment{abstract}
  {\noindent\small 
  {\noindent{\large\textbf\abstractname }
  \thispagestyle{plain}

  }}
  {}
    
\usepackage{fancyhdr}
\fancypagestyle{plain}{
  }

\usepackage{siunitx}
\usepackage{pdfpages}
\usepackage{graphicx}
\usepackage{subcaption}
\usepackage{mathtools}
\usepackage{amsfonts}
\usepackage{cancel}
\usepackage{enumitem}
\usepackage{dsfont}
\usepackage{amssymb}
\usepackage{amsmath}
\usepackage{amsthm}
\usepackage{placeins}
\usepackage{upgreek}
\usepackage{accents}
\usepackage{lipsum}

\newcommand{\bs}{\boldsymbol}

\newcommand{\br}[1]{\left\langle#1\right|}
\newcommand{\ke}[1]{\left|#1\right\rangle}

\newcommand{\commentthis}[1]{}

\author{Sebastian Lahs$^*$, Daniel Comparat}
\title{Polarizabilities as Probes for P, T, and PT Violation} 

\graphicspath{{pics/}} 

\begin{document}

\maketitle
\abstract{\noindent Searches for violations of the fundamental symmetries of parity $P$ and time reversal $T$ in atomic and molecular systems provide a powerful tool for precise measurements of the physics of and beyond the standard model. In this work, we investigate how these symmetry violations affect the response of atoms and molecules to applied electric and magnetic fields. We recover well-known observables such as the $P$ -odd, $T$ -odd spin-electric field coupling that is used for searches of the electron electric dipole moment (eEDM) or the effect of $P$-odd, $T$-even optical rotation in atomic gases. Besides these, we obtain several other possible observables. This includes, in particular, effects that can only be seen when using oscillating or inhomogeneous fields.}
\section{Introduction}\label{sec:Introduction}
\blfootnote{$^*$sebastian.lahs@cnrs.fr}At the present day, precision measurements in atomic and molecular systems are among the most powerful tools to test the standard model of particle physics (SM) at low energy scales \cite{safronova2018search, safronova2023searches, cong2024spin}. This is especially driven by the progress in quantum sensing \cite{jackson2023probing, ye2024essay}. The sensitivity to beyond SM physics enables searches for new particles or interactions that are proposed to answer some of the open questions in the foundations of physics. The interaction of atoms and molecules with these new particles would lead to small shifts in their spectra. The magnitude of almost all of these small corrections is far below the uncertainty at which absolute energy levels can presently be calculated. To avoid this problem, it makes sense to search for observables that are forbidden by the symmetries of atomic physics. If we detect non-zero values of those, their size would directly correspond to some new or exotic interaction. Two such symmetries are the ones of parity $P$ and time reversal $T$. \\
$P$-violating processes were previously used to prove the validity of the Glashow-Weinberg-Salam theory of the electroweak interaction at atomic energy scales \cite{safronova2023searches}. Further, numerous searches for simultaneous violations of $P$ and $T$ were performed in the context of electric dipole moment (EDM) measurements \cite{chupp2019electric, roussy2023improved}. Only few direct searches for $T$, without $P$ violation ($T$-odd $P$-even, also called ToPe) have been performed so far \cite{hopkinson2002interferometric, akdag2023c}.\\

Following previous works \cite{bouchiat1975parity,feinberg1977parity,moskalev1986some,sandars1993p,baryshevsky1994p,derevianko2010cp}, we derive possible observables for measurements of $P$, $T$, and $PT$ -violations in the interaction of electromagnetic fields with atoms and molecules. We try to
be as general as possible by making minimal assumptions on the atoms/molecules under consideration, the spatial and temporal shape of the electromagnetic fields, and the explicit form of the $P$- and $T$- odd potentials. \\

We start by specifying the requirements of the system (sec. \ref{sec: Requirements on the system}) and then introduce perturbation through electric and magnetic fields under a low-order multipole approximation. On this basis, we derive the time-dependent perturbative expressions for the induced electric and magnetic dipole moments and the electric quadrupole moment (sec. \ref{sec: Derivation of the P and T odd polarizabilities}). We separate these terms into irreducible tensor components and analyze their behavior under $P$ and $T$ transformations. This leads us to general expressions for the possible (generalized) atomic polarizabilities. In sec. \ref{sec: resulting observables}, we match some of these polarizabilities to observables that have been considered in previous works. We further identify terms that have not been explored so far. This includes, in particular, interactions with inhomogeneous (sec. \ref{sec: Static inhomogeneous fields}) and time-dependent (sec. \ref{sec: Time-varying fields}) fields. We show that the former leads to observables sensible to all three kinds of symmetry violations. In sec. \ref{sec: Effects from the natural linewidth} we argue that the latter introduces some difficulties in the context of $T$-odd observables.

\section{Derivation of the $P$ and $T$ odd polarizabilities}
\label{sec: Derivation of the P and T odd polarizabilities}
\subsection{Requirements on the system}\label{sec: Requirements on the system}
The atomic or molecular system under consideration has to fulfill a few requirements for the following treatment to be valid. Namely, the Hamiltonian $H^0$ of the unperturbed system needs to commute with parity $P$, the square of the total angular momentum $\bs{J}^2$ and its projection on an arbitrary axis $J_z$. Therefore, we require $H^0$ to be invariant under $SU(2)$ rotations and spatial reflections. Generally, this will be true for any free atom or molecule due to the rotational symmetry of their wavefunctions. Interestingly enough, this holds regardless of how precise our model for the system is. It even remains valid for a full relativistic QED+QCD description of all its inner processes. Only the inclusion of the weak force or hypothetical new physics interactions breaks these symmetries. Therefore, we exclude these from the definition of $H^0$ and instead add them later as small perturbations.
From all this, we know that the eigenstates to $H^0$ can (without approximation) be written in a basis $\ke{n}=\ke{a J m}$. Here, $J$ is the quantum number of the total angular momentum, $m$ is the quantum number related to its projection on a chosen quantization axis, and $a$ includes all the other quantum numbers of the system.\\

This comes with a small caveat. To use all this, we need to be able to prepare a system in these "true" eigenstates. This will, for example, not be possible for chiral molecules whose states of different chirality are separated by such a large energy barrier that they will realistically never be found in a Parity eigenstate (Hund's Paradox \cite{hund1927deutung}). Apart from chiral molecules \cite{condon1937theories,barron2001time},
spontaneous breakings of $P$ and $T$-symmetries can also arise in solid-state systems depending on their crystal symmetries \cite{aroyo2013international,schmid2008some}. These do not originate from the fundamental physics processes we want to consider here. We, therefore, require that all such spontaneous symmetry-breakings are absent. This is generally fulfilled in free atoms and (small) molecules.\\

Throughout this paper, we are using natural units: $c=\hbar=\varepsilon_0=1$. In the following, we refer to the situation where an operator is both odd under $P$ and $T$ as $PT$-violation. We use the notation $\bs{J}$ for the total angular momentum, including possible contributions of the nuclear angular momentum, even though the total angular momentum in such a case would usually be denoted by $\bs{F}$.

\subsection{Perturbative treatment of the interaction with the electromagnetic field}\label{sec: perturbative treatment}
We want to study how the system $H^0$ behaves under the influence of applied electromagnetic fields. To do this, we are using a perturbative treatment in which we add the semiclassical interaction with an electromagnetic field in the 2nd-order multipole approximation:
\begin{align}
 V^{EM}= -\bs{d}\cdot\bs{\mathcal{E}} - \bs{\mu}\cdot\bs{\mathcal{H}} - Q_{ij}\partial_i\mathcal{E}_j\label{eq:V^EM}
\end{align}
where $\bs{d}, \bs{\mu}$, and $\bs{Q}$ are the electric dipole, the magnetic dipole, and the electric quadrupole (Hermitian) operators, respectively. We recall that in a simple single electron system, these moments are defined as follows \cite{steck2007quantum}:
\begin{align*}
 \bs{d}&=e\bs{r}-\mu_B(\bs{\sigma}\times\bs{p})\nonumber\\
 \bs{\mu}&=\mu_B\left(g_s\bs{\sigma}+ \bs{r}\times \bs{p}\right)\nonumber\\
 Q_{ij}&=\frac{e}{2}\left(r_ir_j-\frac{r^2}{3}\delta_{ij} \right)
\end{align*}
Here, $e$ is the absolute value of the electron charge, $\bs{r}$ is the position- and $\bs{p}$ the momentum-vector of the electron, $\mu_B$ is the Bohr magneton, $g_s$, the electron $g$-factor, and $\bs{\sigma}$ is the vector of the three Pauli matrices. In the definition of $\bs{d}$, we included the first-order relativistic correction \cite{bethe2012quantum}. In the remainder of this work, we do not refer to the specific form of these operators and only use their behaviors under parity and time-reversal transformations (see table \ref{tab:my_table}). That way, the treatment remains valid for multi-electron atoms and molecules in which $\bs{d}$, $\bs{\mu}$, and $Q_{ij}$ are obtained by summing over all bodies in the system.\\

As usual, is the energy of the system $H^0+V^{EM}$ is given by:
\begin{align*}
 E=&\br{N}H^0+V^{EM}\ke{N}\nonumber\\
 =&\br{N}H^0\ke{N}-\br{N}\bs{d}\ke{N}\cdot\bs{\mathcal{E}}-\br{N}\bs{\mu}\ke{N}\cdot\bs{\mathcal{H}}-\br{N}Q_{ij}\ke{N}\partial_i\mathcal{E}_j
\end{align*}
with $\ke{N}$ being the eigenvectors of $H=H^0+V^{EM}$.
In the following, we want to treat $V^{EM}$ as a perturbation to $H^0$. We will perform this perturbation explicitly for the electric dipole operator $\bs{d}$. The perturbations for $\bs{\mu}$ and $Q_{ij}$ follow analogously. In appendix \ref{sec: interaction with electromagnetic fields}, we demonstrate how electric and magnetic fields $\bs{\mathcal{E}}$, $\bs{\mathcal{H}}$ can be expressed in regard to the electromagnetic potentials $\phi$ and $\bs{\mathcal{A}}$ as a sum of a static and a time-varying component. Following this, the general perturbative expression, valid for both static and varying fields, can be built from the solutions to the time-independent and dependent problems. Thus, we first perform a time-independent perturbation calculation for the static components of the field and, afterward, a time-dependent calculation for the time-varying components. There are multiple, slightly different, formulations for the \textit{Dirac-type} time-dependent perturbation theory. One has to be careful in choosing the right one to avoid the appearance of secular terms and obtain a correct normalization and convergence in a static limit \cite{langhoff1972aspects,bhattacharya1984modification,bhattacharyya1986perturbative,mandal2012adiabatic}. \\

A free atom (or molecule) is fully degenerate in the quantum number $m$. We assume that there are no further (accidental) degeneracies present. For these assumptions, the states $\ke{n}=\ke{a J m}$ will be "good states"\cite{griffiths2018introduction} in respect to all the operators involved, which allows us to perform a treatment analogous to a nondegenerate one. The maximal field strengths to which such a perturbative treatment will stay valid are highly dependent on the system and states under consideration.

\subsubsection{Time-independent perturbation} \label{sec: Time independent perturbation}

In a static system, $\bs{\mathcal{E}}=\bs{\mathcal{E}}^0$, and $\bs{\mathcal{H}}=\bs{\mathcal{H}}^0$. The 1st order (Rayleigh-Schrödinger) perturbation theory for a potential $V$ leads to 
\begin{align}
 \ke{N} \simeq \ke{n}-\sum_k \frac{\br{k}V\ke{n}}{\omega_{kn}}\ke{k} \label{first order}
\end{align}
with $\omega_{kn}=E_k-E_n=E_{a'J'}-E_{aJ}$. Thus, up to 1st order, the matrix element $\br{N}d_i\ke{N}$ under the influence of the perturbation $V^{EM}$ can be written as:
\begin{align}
 \br{N}d_i\ke{N} &\simeq \br{n}d_i\ke{n}+2 Re\sum_k\Bigg[ \frac{\br{n}d_i\ke{k}\br{k}d_j\ke{n}}{\omega_{kn}}\mathcal{E}_j^0\label{eq: d first order}\\
 &+\frac{\br{n}d_i\ke{k}\br{k}\mu_j\ke{n}}{\omega_{kn}}\mathcal{H}_j^0+\frac{\br{n}d_i\ke{k}\br{k}Q_{j\varkappa} \ke{n}}{\omega_{kn}}\partial_j\mathcal{E}_\varkappa^0 \Bigg]\nonumber
\end{align}

\subsubsection{Time-dependent perturbation} 
\label{sec: Time dependent perturbation}

For deriving the time-dependent contributions, we start by writing the non static part of Eq. (\ref{eq:V^EM}) in respect to the electromagnetic vector potential $\bs{\mathcal{A}}$ (see Eqs. (\ref{eq: E E0 Et}), (\ref{eq: H H0 Ht}), and (\ref{eq: E H derivatives})):
\begin{align*}
 V^{EM}(t) &\equiv\sum_{\ell\neq 0} V^\ell=\sum_{\ell\neq 0}\Big[-\bs{d}\cdot\bs{\mathcal{E}}^\ell - \bs{\mu}\cdot\bs{\mathcal{H}}^\ell - Q_{ij}\partial_i\mathcal{E}^\ell\Big]\\
 &=\sum_{\ell\neq 0}2Re\left(\left[-i\omega^\ell\bs{d}\cdot\bs{\mathcal{A}}^\ell -i\bs{\mu}\cdot\big(\bs{k}^\ell\times\bs{\mathcal{A}}^\ell\big) + \omega^\ell Q_{ij} k_i^\ell \mathcal{A}_j^\ell \right] e^{-i\omega^\ell t}\right)\\
 & \equiv \sum_{\ell\neq 0}\Big[F^\ell e^{-i\omega^\ell t} +(F^\ell)^
 \dagger e^{i\omega^\ell t}\Big]
\end{align*}
We can see that we have the special situation of a perturbation periodic in time (see, for example \cite{langhoff1972aspects,landau2013quantum6}). In this case, the induced dipole moment is given by:
\begin{align}
\br{N}d_i\ke{N}(t)\simeq \br{n}&d_i\ke{n}\label{eq:NdtM}\\
-\sum_{\ell\neq 0}\sum_{k}&\left[\frac{\br{n}d_i\ke{k} \br{k}F^\ell\ke{n}}{\omega_{k n}-\omega^\ell}+\frac{\br{n}F^\ell\ke{k}\br{k}d_i\ke{n} }{\omega_{k n}+\omega^\ell}\right] e^{-i \omega^\ell t}\nonumber\\
+&\left[\frac{\br{n}d_i\ke{k} \br{k}(F^\ell)^\dagger\ke{n}}{\omega_{k n}+\omega^\ell}+\frac{\br{n}(F^\ell)^\dagger\ke{k}\br{k}d_i\ke{n} }{\omega_{k n}-\omega^\ell}\right] e^{i \omega^\ell t}\nonumber
\end{align}
This expression is only true far from resonance $\left(|\omega^\ell|\approx|\omega_{kn}|\right)$. The situation close to resonance is discussed in sec. \ref{sec: Effects from the natural linewidth}.\\
By expanding the fractions and identifying $-i\omega^\ell F^\ell e^{-i \omega^\ell t}+i\omega^\ell (F^\ell)^\dagger e^{i \omega^\ell t}$ with the time derivative of $V^{\ell}$, we can rewrite this expression as:
\begin{align}
 \br{n}&d_i\ke{n} -2Re\sum_{\ell\neq 0}\sum_k\frac{\br{n}d_i\ke{k}\br{k}\big[\omega_{kn}V^{\ell}+i\dot{V}^{\ell}\big]\ke{n}}{\omega_{k n}^2-(\omega^\ell)^2}
 \label{eq:NdtM2}
\end{align}
In the static case, $\omega^\ell=0$, $\dot{V}^{EM}$ vanishes, and the equation returns to the same form as Eq. (\ref{eq: d first order}). 
\subsubsection{General perturbation}
We can now combine both the expression for purely static and purely time-dependent fields. After reinserting the definition of $V^\ell$, we arrive at a general expression for the electric dipole moment element:
\begin{align}
\langle d_i\rangle\simeq \br{n}d_i\ke{n}
+Re 
 &\Big[\prescript{dd}{}{\alpha}_{ij}^{\ell }\Big]\mathcal{E}_{j}^\ell+Im\Big[\prescript{dd}{}{\alpha'}_{ij}^{\ell }\Big]\dot{\mathcal{E}}_{j}^\ell\label{eq:NdN}\\
 +Re 
 &\Big[\prescript{d\mu}{}{\alpha}_{ij}^{\ell }\Big]\mathcal{H}_{j}^\ell+Im\Big[\prescript{d\mu}{}{\alpha'}_{ij}^{\ell }\Big]\dot{\mathcal{H}}_{j}^\ell\nonumber\\
 +Re 
 &\Big[\prescript{dQ}{}{\alpha}_{ij\varkappa}^{\ell }\Big]\partial_j\mathcal{E}_{\varkappa}^{\ell}+Im\Big[\prescript{dQ}{}{\alpha'}_{ij\varkappa}^{\ell }\Big]\partial_j\dot{\mathcal{E}}_{\varkappa}^\ell\nonumber
\end{align}
Here and in the following, we imply, according to the Einstein convention, the summation over $\ell$ (including the static case $\ell=0$, $\omega^0=0$). Above, we further introduced the (generalized) polarizabilities:
\begin{align}
\prescript{d\mu}{}{\alpha}_{ij}^{\ell }&=2\sum_k\frac{\br{n}d_i\ke{k}\br{k}\mu_j\ke{n}}{\omega_{kn}^2-(\omega^\ell)^2}\omega_{kn} \label{eq: def alpha}\\
\prescript{d\mu}{}{\alpha'}_{ij}^{\ell }&=-2\sum_k\frac{\br{n}d_i\ke{k}\br{k}\mu_j\ke{n}}{\omega_{kn}^2-(\omega^\ell)^2} \nonumber
\end{align}
with $\prescript{d d}{}{\alpha}_{ij}^{\ell}$, $\prescript{d d}{}{\alpha'}_{ij}^{\ell}$, $\prescript{d Q}{}{\alpha}_{ij\varkappa}^{\ell}$, and $\prescript{d Q}{}{\alpha'}_{ij\varkappa}^{\ell}$ following analogously.\\\\
Likewise, we can perform the same derivation for the magnetic dipole moment, which leads to:
\begin{align}
\langle \mu_i\rangle \simeq \br{n}\mu_i\ke{n}
+Re 
 &\Big[\prescript{\mu d}{}{\alpha}_{ij}^{\ell }\Big]\mathcal{E}_{j}^\ell+Im\Big[\prescript{\mu d}{}{\alpha'}_{ij}^{\ell }\Big]\dot{\mathcal{E}}_{j}^\ell\label{eq:NmuN}\\
 +Re 
 &\Big[\prescript{\mu\mu}{}{\alpha}_{ij}^{\ell }\Big]\mathcal{H}_{j}^\ell+Im\Big[\prescript{\mu\mu}{}{\alpha'}_{ij}^{\ell }\Big]\dot{\mathcal{H}}_{j}^\ell\nonumber\\
 +Re 
 &\Big[\prescript{\mu Q}{}{\alpha}_{ij\varkappa}^{\ell }\Big]\partial_j\mathcal{E}_{\varkappa}^{\ell}+Im\Big[\prescript{\mu Q}{}{\alpha'}_{ij\varkappa}^{\ell }\Big]\partial_j\dot{\mathcal{E}}_{\varkappa}^\ell\nonumber
\end{align}
Finally, for the electric quadrupole moment, we get:
\begin{align}
\langle Q_{ij} \rangle \simeq \br{n}Q_{ij}\ke{n}
+Re 
 &\Big[\prescript{Qd}{}{\alpha}_{ij\varkappa}^{\ell }\Big]\mathcal{E}_{\varkappa}^\ell+Im\Big[\prescript{Qd}{}{\alpha'}_{ij}^{\ell }\Big]\dot{\mathcal{E}}_{\varkappa}^\ell\label{eq:NQN}\\
 +Re 
 &\Big[\prescript{Q\mu}{}{\alpha}_{ij\varkappa}^{\ell }\Big]\mathcal{H}_{\varkappa}^\ell+Im\Big[\prescript{Q\mu}{}{\alpha'}_{ij\varkappa}^{\ell }\Big]\dot{\mathcal{H}}_{\varkappa}^\ell\nonumber\\
 +Re 
 &\Big[\prescript{QQ}{}{\alpha}_{ij\varkappa l}^{\ell }\Big]\partial_\varkappa\mathcal{E}_{l}^{\ell}+Im\Big[\prescript{dQ}{}{\alpha'}_{ij\varkappa l}^{\ell }\Big]\partial_\varkappa\dot{\mathcal{E}}_{l}^\ell\nonumber
\end{align}
 For simplicity and consistency, we will refer to all $\alpha$ terms as (generalized) polarizabilities, even if, for specific cases, other names such as 
 \textit{permeability} or \textit{quadrupolarizability} exist. We also refer to the field-independent static moments as polarizabilities.

\subsection{Violation of discrete symmetries}
In usual atomic physics derivations for electric and magnetic dipole moments, most of the terms in Eqs. (\ref{eq:NdN}), (\ref{eq:NmuN}), and (\ref{eq:NQN}) are absent. As we will see in a moment, this is due to the symmetries of $H^0$. These terms can, however, become observable when $H^0$ is additionally perturbed by $P$- and $T$-odd potentials $V^N$:
\begin{align}
 H&=H^0+V^{EM}+V^N \label{eq: full hamiltonian}\\
 V^N&=-V^P-V^{PT}-V^T\nonumber
\end{align}
$V^P$ is an internal $P$-odd, $T$-even, $V^{PT}$ a $P$-odd, $T$-odd, and $V^T$ a $P$-even, $T$-odd interaction. The global minus sign in $V^N$ does not carry a physical meaning and is chosen in analogy to Eq (\ref{eq:V^EM}).\\
In the following, we will take $\ke{N}$ to be the eigenstates of this full Hamiltonian $H$.
To make the treatment more general, we do not specify the structure of these operators but instead only use their transformation behaviors under $P$ and $T$. For explicit examples of $V^P$, $V^{PT}$, and $V^T$ operators that arise in electron-nucleon interactions, see, for instance, Ref. \cite{khriplovich1991parity,khriplovich2012cp}. We restrict ourselves to symmetry-violating interactions that take place inside the free atom or molecule and do not directly couple to the externally applied fields. For many situations, this assumption makes sense as the internal electric and magnetic fields will exceed the externally applied ones in strength. \\

In the following, we investigate which terms in Eqs. (\ref{eq:NdN}), (\ref{eq:NmuN}), and (\ref{eq:NQN}) correspond to the breakings of which particular symmetries. The most important properties of the $P$ and $T$ operators are indicated in table \ref{tab:my_table} and appendix \ref{sec: The time-reversal operator T}.
\begin{figure}[t]
\begin{minipage}{0.49\textwidth}
\centering
\begin{tabular}{l|cccc|ccc}
 & $\bs{d}$& $\bs{\mu}$& $\bs{Q}$&$\bs{J}$ & $V^P$& $V^{PT}$&$V^{T}$\\ \hline
$\bs{P}$& - & + & +&+ & -& -&+\\
$\bs{T}$& + & - & +&- & +& -&-\\\end{tabular}
\end{minipage}
\hfill
\begin{minipage}{0.49\textwidth}
\begin{align*}
 \quad P\ke{a J m}&=\pm\ke{a J m}\\
 \quad T\ke{a J m}&=(-1)^{J-m} \ke{a J -m}
\end{align*}
\end{minipage}
\captionof{table}{Transformation properties of the three multipole operators, the total angular momentum, the symmetry violating potentials, and the free Hamiltonian eigenstates $\ke{n}=\ke{a J m}$ under the linear parity $P$, and antilinear time reversal $T$ operator. They satisfy $PP=T^\dagger T=1$.}
\label{tab:my_table}
\end{figure}
\subsection{$P$ and $T$ odd polarizabilities }\label{sec: P and T odd polarizabilities}
The effects of $V^{P}$, $V^{T}$, and $V^{PT}$ on $\langle \bs{d} \rangle$, $\langle \bs{\mu} \rangle$, and $\langle {Q}_{ij} \rangle$ can again be derived using perturbation theory. We discuss the details of this procedure in appendix \ref{sec: Simultaneous perturbation through electromagnetic an symmetry violating potentials}. From it, we get a new, more general, expression for the induced electric dipole moment $\langle d_i \rangle$ in terms of $P,\;T$-even and -odd polarizabilities:
\begin{align}
\langle d_i\rangle \simeq Re&\Big[\prescript{d}{}{\alpha}_{i}^{}+\prescript{d}{P}{\alpha}_{i}^{}+\prescript{d}{PT}{\alpha}_{i}^{}+\prescript{d}{T}{\alpha}_{i}^{}\Big]\label{eq: <d> P PT T}\\
+Re 
 &\Big[\prescript{dd}{}{\alpha}_{ij}^{\ell}+\prescript{dd}{P}{\alpha}_{ij}^{\ell}+\prescript{dd}{PT}{\alpha}_{ij}^{\ell}+\prescript{dd}{T}{\alpha}_{ij}^{\ell}\Big]\mathcal{E}_{j}^\ell\nonumber\\
+Re&\left[\prescript{d \mu}{}{\alpha}_{ij}^{\ell}+\prescript{d \mu}{P}{\alpha}_{ij}^{\ell}+\prescript{d \mu}{PT}{\alpha}_{ij}^{\ell}+\prescript{d \mu}{T}{\alpha}_{ij}^{\ell}\right]\mathcal{H}_{j}^\ell\nonumber\\
+Re&\left[\prescript{d Q}{}{\alpha}_{ij}^{\ell}+\prescript{d Q}{P}{\alpha}_{ij}^{\ell}+\prescript{d Q}{PT}{\alpha}_{ij}^{\ell}+\prescript{dQ}{T}{\alpha}_{ij}^{\ell}\right]\partial_j\mathcal{E}_{\varkappa}^{\ell} \nonumber\\
+Im&\Big[\prescript{dd}{}{\alpha'}_{ij}^{\ell}+\prescript{dd}{P}{\alpha'}_{ij}^{\ell}+\prescript{dd}{PT}{\alpha'}_{ij}^{\ell}+\prescript{dd}{T}{\alpha'}_{ij}^{\ell}\Big]\dot{\mathcal{E}}_{j}^\ell\nonumber\\
+Im&\left[\prescript{d \mu}{}{\alpha'}_{ij}^{\ell}+\prescript{d \mu}{P}{\alpha'}_{ij}^{\ell}+\prescript{d \mu}{PT}{\alpha'}_{ij}^{\ell}+\prescript{d \mu}{T}{\alpha'}_{ij}^{\ell}\right]\dot{\mathcal{H}}_{j}^\ell\nonumber\\
+Im&\left[\prescript{d Q}{}{\alpha'}_{ij}^{\ell}+\prescript{d Q}{P}{\alpha'}_{ij}^{\ell}+\prescript{d Q}{PT}{\alpha'}_{ij}^{\ell}+\prescript{dQ}{T}{\alpha'}_{ij}^{\ell}\right]\partial_j\dot{\mathcal{E}}_{\varkappa}^\ell \nonumber
\end{align}
with $\prescript{d}{}{\alpha}_{i}^{}=\br{n}d_i\ke{n}$. The first term in each row is, as before, defined according to eq: (\ref{eq: def alpha}). As indicated by their lower left indices, the remaining polarizabilities are proportional to one of the three potentials $V^{P}$, $V^{T}$, and $V^{PT}$. We define $\prescript{d}{PT}{\alpha}_{i}^{}$ as (see sec. \ref{sec: 1st order in V^EM}):
\begin{align}
 \prescript{d}{PT}{\alpha}_{i}^{}= 2 \sum_k \frac{\br{n}d_i\ke{k}\br{k}V^{PT}\ke{n}}{\omega_{kn}} \label{eq: alpha d PT}
\end{align}
with $\prescript{d}{P}{\alpha}_{i}^{}$ and $\prescript{d}{T}{\alpha}_{i}^{}$ following accordingly.\\
The remaining terms are defined analogously to:
\begin{align}
\prescript{d \mu}{PT}{\alpha}_{ij}^{\ell}= 2 \sum_{k,\,l}\Bigg[ &\frac{\br{n}V^{PT}\ke{l}\br{l}d_i\ke{k}\br{k}\mu_j\ke{n}}{(\omega_{kn}^2-(\omega^\ell)^2)\omega_{ln}}\omega_{kn}+\frac{\br{n}d_i\ke{k}\br{k}\mu_j\ke{l}\br{l}V^{PT}\ke{n}}{(\omega_{kn}^2-(\omega^\ell)^2)\omega_{ln}}\omega_{kn}\nonumber\\
  +&\frac{\br{n}d_i\ke{k}\br{k}V^{PT}\ke{l}\br{l}\mu_j\ke{n}}{(\omega_{kn}^2-(\omega^\ell)^2)(\omega_{ln}^2-(\omega^\ell)^2)}\left(\omega_{kn}\omega_{ln}+(\omega^\ell)^2\right)\Bigg]\nonumber\\
  \prescript{d \mu}{PT}{\alpha'}_{ij}^{\ell}=-2\sum_{k,\,l}\Bigg[ &\frac{\br{n}V^{PT}\ke{l}\br{l}d_i\ke{k}\br{k}\mu_j\ke{n}}{(\omega_{kn}^2-(\omega^\ell)^2)\omega_{ln}}+\frac{\br{n}d_i\ke{k}\br{k}\mu_j\ke{l}\br{l}V^{PT}\ke{n}}{(\omega_{kn}^2-(\omega^\ell)^2)\omega_{ln}}\nonumber\\
  +&\frac{\br{n}d_i\ke{k}\br{k}V^{PT}\ke{l}\br{l}\mu_j\ke{n}}{(\omega_{kn}^2-(\omega^\ell)^2)(\omega_{ln}^2-(\omega^\ell)^2)}\left(\omega_{ln}+\omega_{kn}\right)\Bigg]\label{eq: alpha PT d mu}
\end{align}
Likewise, expressions for $\langle\mu_i\rangle$ and $\langle Q_{ij}\rangle$ in terms of $P$ and $T$ odd polarizabilities can be derived.
\subsection{Full expressions of the induced moments}\label{sec: Full expressions of the induced moments}
The form of $\langle d_i\rangle$ in Eq. (\ref{eq: <d> P PT T}), where the induced moment is related to polarizability tensors, is not very practicable yet. To arrive at useful expressions, we first need to decompose these tensors into their irreducible components. After this, we can analyze these components concerning their transformation behavior under parity $P$ and time reversal $T$. We present the procedure of how to reduce and analyze all polarizability terms belonging to $\langle d_i\rangle$, $\langle \mu_i\rangle$, and $\langle Q_{ij}\rangle$ in appendix \ref{app: Analyzing the polarizabilities in terms of parity and time reversal}. \\

For example, for $\prescript{d \mu}{PT}{\alpha}_{ij}^{\ell}$ angular momentum algebra together with the analysis of its behavior under $T$, lets us express the tensor in respect to three real constants $\prescript{d\mu}{PT}{\alpha}_s^{\ell }$, $\prescript{d\mu}{PT}{\alpha}_v^{\ell }$, and $\prescript{d\mu}{PT}{\alpha}_t^{\ell }$ (appendix \ref{sec: Behavior of the d mu terms under time reversal}):
\begin{align*}
\prescript{d\mu}{PT}{\alpha}_{ij}^{\ell }=\delta_{ij}\; \prescript{d\mu}{PT}{\alpha}_s^{\ell }+i\varepsilon_{ij\varkappa}\mathcal{J}_\varkappa\; \prescript{d\mu}{PT}{\alpha}_v^{\ell }+\mathcal{Q}^{\bs{J}^2}_{ij}\; \prescript{d\mu}{PT}{\alpha}_t^{\ell }
\end{align*}
where $\delta_{ij}$ is the Kronecker delta and $\varepsilon_{ij\varkappa}$ is the Levi-Civita tensor. $\bs{J}$ is the (conserved) total angular momentum of the system, and $\bs{\mathcal{J}}=\br{n}\bs{J}\ke{n}$. $\bs{\mathcal{Q}}^{\bs{J}^2}$ is a symmetric traceless matrix defined by $\mathcal{Q}^{\bs{J}^2}_{ij}=\br{n}\tfrac{1}{2}\left[J_i J_j+J_j J_i-\frac{2}{3}\delta_{ij}\sum_lJ_l J_l\right]\ke{n}$. The induced electric dipole moment $\langle d_i \rangle$ in Eq. (\ref{eq: <d> P PT T}) depends solely on the real part of $\prescript{d \mu}{PT}{\alpha}_{ij}^{\ell}$. The term $i\varepsilon_{ij\varkappa}\mathcal{J}_\varkappa\; \prescript{d\mu}{PT}{\alpha}_v^{\ell }$ can, therefore, not contribute.\\

Such an analysis for all the terms in Eq. (\ref{eq: <d> P PT T}), leads us to a general expression for the induced electric dipole moment:
\begin{align}
 \langle \bs{d} \rangle &\simeq \prescript{d}{PT}{\alpha}_{}^{}\, \bs{\mathcal{J}}+\prescript{dd}{}{\alpha}_{s}^{\ell}\, \bs{\mathcal{E}}^\ell+\prescript{dd}{}{\alpha'}_{v}^{\ell}\, (\dot{\bs{\mathcal{E}}}^\ell\times\bs{\mathcal{J}})+\prescript{dd}{}{\alpha}_{t}^{\ell}\, \bs{\mathcal{Q}}^{\bs{J}^2}\cdot\bs{\mathcal{E}}^\ell\label{eq: <d>}\\
 &+\prescript{d\mu}{P}{\alpha'}_{s}^{\ell}\, \dot{\bs{\mathcal{H}}}^\ell+\prescript{d\mu}{P}{\alpha}_{v}^{\ell}\, ({\bs{\mathcal{H}}}^\ell\times\bs{\mathcal{J}})+\prescript{d\mu}{P}{\alpha'}_{t}^{\ell}\, \bs{\mathcal{Q}}^{\bs{J}^2}\cdot\dot{\bs{\mathcal{H}}}^\ell\nonumber\\
 &+\prescript{d\mu}{PT}{\alpha}_{s}^{\ell}\, \bs{\mathcal{H}}^\ell+\prescript{d\mu}{PT}{\alpha'}_{v}^{\ell}\, (\dot{\bs{\mathcal{H}}}^\ell\times\bs{\mathcal{J}})+\prescript{d\mu}{PT}{\alpha}_{t}^{\ell}\, \bs{\mathcal{Q}}^{\bs{J}^2}\cdot\bs{\mathcal{H}}^\ell\nonumber\\
 &+\left(\prescript{dQ}{P}{\alpha'}_{t_s}^{\ell}\,\mathcal{S}_{ij\varkappa}^{\bs{J^3}}+\prescript{dQ}{P}{\alpha'}_{v_m}^{\ell}\,\mathcal{M}_{ij\varkappa}^{\bs{J}}\right)\hat{\bs{e}}_i \partial_j \dot{\mathcal{E}}^\ell_\varkappa+\prescript{dQ}{P}{\alpha}_{t_m}^{\ell}\,\mathcal{M}_{ij\varkappa}^{\bs{J}^2}\hat{\bs{e}}_i \partial_j {\mathcal{E}}^\ell_\varkappa\nonumber\\
 &+\left(\prescript{dQ}{PT}{\alpha}_{t_s}^{\ell}\,\mathcal{S}_{ij\varkappa}^{\bs{J}^3}+\prescript{dQ}{PT}{\alpha}_{v_m}^{\ell}\,\mathcal{M}_{ij\varkappa}^{\bs{J}}\right)\hat{\bs{e}}_i \partial_j {\mathcal{E}}^\ell_\varkappa+\prescript{dQ}{PT}{\alpha'}_{t_m}^{\ell}\,\mathcal{M}_{ij\varkappa}^{\bs{J}^2}\hat{\bs{e}}_i \partial_j \dot{\mathcal{E}}^\ell_\varkappa\nonumber
\end{align}
We recall that $\bs{\mathcal{E}}^\ell$ is the electric and $\bs{\mathcal{H}}^\ell$ the magnetic field, while $\dot{\bs{\mathcal{E}}^\ell}$ and $\dot{\bs{\mathcal{H}}^\ell}$ are their time derivatives. $\mathcal{S}_{ij\varkappa}^{\bs{J}^3}$, $ \mathcal{M}_{ij\varkappa}^{\bs{J}}$, and $\mathcal{M}_{ij\varkappa}^{\bs{J}^2}$ are rank-3 tensors that are defined in sec. \ref{sec:electric dipole- electric quadrupole}. $\hat{\bs{e}}_i$ is a unit vector along the (Cartesian) coordinate $i$. \\
The real coefficients $\alpha$, $\alpha'$ are the (generalized) polarizabilities that result from the tensor decomposition of the quantities defined in Eqs. (\ref{eq: alpha d PT}), (\ref{eq: alpha PT d mu}).\\
The lower right index of the $\alpha$ defines which structure they belong to ($s$=scalar trace, $v$=antisymmetric vector, $t$=symmetric traceless quadrupole tensor, $t_s$= symmetric traceless octupole tensor, $v_m$=mixed symmetry vector, $t_m$=mixed symmetry tensor ).\\
The lower left index of the $\alpha$ tells if the interaction requires perturbation by a $P$, $T$, or both $PT$ odd potential. \\

Equivalently to the electric dipole, we can find the induced magnetic dipole moment:
\begin{align}
 \langle \bs{\mu} \rangle &\simeq \prescript{\mu}{}{\alpha}_{}^{}\, \bs{\mathcal{J}}+\prescript{\mu\mu}{}{\alpha}_{s}^{\ell}\, \bs{\mathcal{H}}^\ell+\prescript{\mu\mu}{}{\alpha'}_{v}^{\ell}\, (\dot{\bs{\mathcal{H}}}^\ell\times\bs{\mathcal{J}})+\prescript{\mu\mu}{}{\alpha}_{t}^{\ell}\, \bs{\mathcal{Q}}^{\bs{J}^2}\cdot\bs{\mathcal{H}}^\ell\label{eq: <mu>}\\
 &+\prescript{\mu d}{P}{\alpha'}_{s}^{\ell}\, \dot{\bs{\mathcal{E}}}^\ell+\prescript{\mu d}{P}{\alpha}_{v}^{\ell}\, ({\bs{\mathcal{E}}}^\ell\times\bs{\mathcal{J}})+\prescript{\mu d}{P}{\alpha'}_{t}^{\ell}\, \bs{\mathcal{Q}}^{\bs{J}^2}\cdot\dot{\bs{\mathcal{E}}}^\ell\nonumber\\
 &+\prescript{\mu d}{PT}{\alpha}_{s}^{\ell}\, \bs{\mathcal{E}}^\ell+\prescript{\mu d}{PT}{\alpha'}_{v}^{\ell}\, (\dot{\bs{\mathcal{E}}}^\ell\times\bs{\mathcal{J}})+\prescript{\mu d}{PT}{\alpha}_{t}^{\ell}\, \bs{\mathcal{Q}}^{\bs{J}^2}\cdot\bs{\mathcal{E}}^\ell\nonumber\\
 &+\left(\prescript{\mu Q}{T}{\alpha'}_{t_s}^{\ell}\,\mathcal{S}_{ij\varkappa}^{\bs{J}^3}+\prescript{\mu Q}{T}{\alpha'}_{v_m}^{\ell}\,\mathcal{M}_{ij\varkappa}^{\bs{J}}\right)\hat{\bs{e}}_i \partial_j \dot{\mathcal{E}}^\ell_\varkappa+\prescript{\mu Q}{T}{\alpha}_{t_m}^{\ell}\,\mathcal{M}_{ij\varkappa}^{\bs{J}^2}\hat{\bs{e}}_i \partial_j {\mathcal{E}}^\ell_\varkappa\nonumber\\
 &+\left(\prescript{\mu Q}{}{\alpha}_{t_s}^{\ell}\,\mathcal{S}_{ij\varkappa}^{\bs{J}^3}+\prescript{\mu Q}{}{\alpha}_{v_m}^{\ell}\,\mathcal{M}_{ij\varkappa}^{\bs{J}}\right)\hat{\bs{e}}_i \partial_j {\mathcal{E}}^\ell_\varkappa+\prescript{\mu Q}{}{\alpha'}_{t_m}^{\ell}\,\mathcal{M}_{ij\varkappa}^{\bs{J}^2}\hat{\bs{e}}_i \partial_j \dot{\mathcal{E}}^\ell_\varkappa\nonumber
\end{align}
For the electric quadrupole, we get:
\begin{align}
 \langle Q_{ij}\rangle &\simeq \prescript{Q}{}{\alpha}_{}^{}\,\mathcal{Q}^{\bs{J}^2}_{ij} +\prescript{QQ}{}{\alpha}_{ij\varkappa l}^{\ell}\,\partial_\varkappa \mathcal{E}^\ell_l\label{eq: <Q>}\\
 &+\left(\prescript{Qd}{P}{\alpha'}_{t_s}^{\ell}\,\mathcal{S}_{\varkappa i j}^{\bs{J}^3}+\prescript{Qd}{P}{\alpha'}_{v_m}^{\ell}\,\mathcal{M}_{\varkappa i j}^{\bs{J}}\right) \dot{\mathcal{E}}^\ell_\varkappa+\prescript{Qd}{P}{\alpha}_{t_m}^{\ell}\,\mathcal{M}_{\varkappa i j}^{\bs{J}^2} {\mathcal{E}}^\ell_\varkappa\nonumber\\
 &+\left(\prescript{Qd}{PT}{\alpha}_{t_s}^{\ell}\,\mathcal{S}_{\varkappa i j}^{\bs{J}^3}+\prescript{Qd}{PT}{\alpha}_{v_m}^{\ell}\,\mathcal{M}_{\varkappa i j}^{\bs{J}}\right) {\mathcal{E}}^\ell_\varkappa+\prescript{Qd}{PT}{\alpha'}_{t_m}^{\ell}\,\mathcal{M}_{\varkappa i j}^{\bs{J}^2} \dot{\mathcal{E}}^\ell_\varkappa\nonumber\\
 &+\left(\prescript{Q\mu}{T}{\alpha'}_{t_s}^{\ell}\,\mathcal{S}_{\varkappa i j}^{\bs{J}^3}+\prescript{Qd}{T}{\alpha'}_{v_m}^{\ell}\,\mathcal{M}_{\varkappa i j}^{\bs{J}}\right) \dot{\mathcal{H}}^\ell_\varkappa+\prescript{Q\mu}{T}{\alpha}_{t_m}^{\ell}\,\mathcal{M}_{\varkappa i j}^{\bs{J}^2} {\mathcal{H}}^\ell_\varkappa\nonumber\\
 &+\left(\prescript{Q\mu}{}{\alpha}_{t_s}^{\ell}\,\mathcal{S}_{\varkappa i j}^{\bs{J}^3}+\prescript{Q\mu}{}{\alpha}_{v_m}^{\ell}\,\mathcal{M}_{\varkappa i j}^{\bs{J}}\right) {\mathcal{H}}^\ell_\varkappa+\prescript{Q\mu}{}{\alpha'}_{t_m}^{\ell}\,\mathcal{M}_{\varkappa i j}^{\bs{J}^2} \dot{\mathcal{H}}^\ell_\varkappa\nonumber
\end{align}
We can see that each field configuration of $\bs{\mathcal{E}}^\ell$, $\bs{\mathcal{H}}^\ell$, $\dot{\bs{\mathcal{E}}^\ell}$, $\dot{\bs{\mathcal{H}}^\ell}$, and $\bs{\mathcal{J}}$ uniquely couples to a specific symmetry violation.
\subsection{Induced $\langle V^P\rangle$, $\langle V^{PT}\rangle$, and $\langle V^T\rangle$} 
\label{sec: Induced V^P V^PT V^T}

In addition to the induced moments that are defined above, also the expectation values of the symmetry-violating potentials $\langle V^P\rangle$, $\langle V^{PT}\rangle$, and $\langle V^T\rangle$ themselves contribute to the full perturbative expressions of $\langle H \rangle$, and $\ke{N}$. In appendix \ref{app: Derivation of V^P V^PT V^T} we show how the lowest order contributions to $\langle V^P\rangle$, $\langle V^{PT}\rangle$, and $\langle V^T\rangle$ can be derived. Up to the 1st order in $V^{EM}$, the only nonzero terms are given by:
\begin{align}
 \langle V^P\rangle &\simeq \prescript{d}{P}{\beta'}_{}^{\ell}\,\bs{\mathcal{J}}\cdot\dot{\bs{\mathcal{E}}}^\ell\label{eq: <V>}\\
 \langle V^{PT}\rangle &\simeq\prescript{d}{PT}{\beta}_{}^{\ell}\,\bs{\mathcal{J}}\cdot{\bs{\mathcal{E}}}^\ell\nonumber\\
 \langle V^T\rangle &\simeq \prescript{\mu}{T}{\beta'}_{}^{\ell}\,\bs{\mathcal{J}}\cdot\dot{\bs{\mathcal{H}}}^\ell+\prescript{Q}{T}{\beta'}_{}^{\ell}\,\mathcal{Q}^{\bs{J}^2}_{ij}\partial_i\dot{\mathcal{E}}^\ell_j\nonumber
\end{align}
Here, the $\beta$ are prefactors similar to the polarizabilities $\alpha$. For example:
\begin{align*}
 \langle V^P \rangle \simeq
 -2Im \sum_{k} 
 &\frac{\br{n}V^P\ke{k}\br{k}d_i\ke{n}}{\omega_{kn}^2-(\omega^\ell)^2}\dot{\mathcal{E}}_{i}^\ell \equiv \prescript{d}{P}{\beta'}_{}^{\ell}\,\bs{\mathcal{J}}\cdot\dot{\bs{\mathcal{E}}}^\ell
\end{align*}
Regarding the (time-dependent) mixing of states or energy shifts, the terms above constitute the lowest order (in $V^{EM}$) corrections that violate $P$, $PT$, and $T$, respectively.

\section{Resulting observables}
\label{sec: resulting observables}
We now connect the terms we derived above to possible experimental observables. It will become apparent that the derivation presented in this paper recovers many observables that have been considered in other works. Additionally, we obtain some $P$-, $T$-, and $PT$-odd observables that have, to our knowledge, not been discussed so far.
\subsection{Static homogeneous fields} \label{sec: Static homogeneous fields}
We start by reviewing the situation for static homogeneous fields ($\dot{\mathcal{E}}_i,\, \dot{\mathcal{H}}_i=0,\,\partial_i{\mathcal{E}}_j=0$). Under these considerations, most terms in Eqs. (\ref{eq: <d>}), (\ref{eq: <mu>}), and (\ref{eq: <Q>}) vanish. We are left with:
\begin{align}
 \langle \bs{d} \rangle &\simeq \prescript{d}{PT}{\alpha}_{}\, \bs{\mathcal{J}}+\prescript{dd}{}{\alpha}_{s}\, \bs{\mathcal{E}}+\prescript{dd}{}{\alpha}_{t}\, \bs{\mathcal{Q}}^{\bs{J}^2}\cdot\bs{\mathcal{E}}\label{eq: d mu homo}\\
 &+\prescript{d\mu}{P}{\alpha}_{v}\, ({\bs{\mathcal{H}}}\times\bs{\mathcal{J}})+\prescript{d\mu}{PT}{\alpha}_{s}\, \bs{\mathcal{H}}+\prescript{d\mu}{PT}{\alpha}_{t}\, \bs{\mathcal{Q}}^{\bs{J}^2}\cdot\bs{\mathcal{H}}\nonumber\\
 \langle \bs{\mu} \rangle &\simeq \prescript{\mu}{}{\alpha}_{}\, \bs{\mathcal{J}}+\prescript{\mu\mu}{}{\alpha}_{s}\, \bs{\mathcal{H}}+\prescript{\mu\mu}{}{\alpha}_{t}\, \bs{\mathcal{Q}}^{\bs{J}^2}\cdot\bs{\mathcal{H}}\nonumber\\
 &+\prescript{\mu d}{P}{\alpha}_{v}\, ({\bs{\mathcal{E}}}\times\bs{\mathcal{J}})+\prescript{\mu d}{PT}{\alpha}_{s}\, \bs{\mathcal{E}}+\prescript{\mu d}{PT}{\alpha}_{t}\, \bs{\mathcal{Q}}^{\bs{J}^2}\cdot\bs{\mathcal{E}} \nonumber\\
 \langle Q_{ij}\rangle &\simeq \prescript{Q}{}{\alpha}_{}^{}\,\mathcal{Q}^{\bs{J}^2}_{ij} +\prescript{QQ}{}{\alpha}_{ij\varkappa l}^{}\,\partial_\varkappa \mathcal{E}_l\nonumber\\
 &+\left(\prescript{Qd}{PT}{\alpha}_{t_s}^{}\,\mathcal{S}_{\varkappa i j}^{\bs{J}^3}+\prescript{Qd}{PT}{\alpha}_{v_m}^{}\,\mathcal{M}_{\varkappa i j}^{\bs{J}}\right) {\mathcal{E}}_\varkappa+\prescript{Qd}{P}{\alpha}_{t_m}^{}\,\mathcal{M}_{\varkappa i j}^{\bs{J}^2} {\mathcal{E}}_\varkappa\nonumber\\
 &+\left(\prescript{Q\mu}{}{\alpha}_{t_s}^{}\,\mathcal{S}_{\varkappa i j}^{\bs{J}^3}+\prescript{Q\mu}{}{\alpha}_{v_m}^{}\,\mathcal{M}_{\varkappa i j}^{\bs{J}}\right) {\mathcal{H}}_\varkappa+\prescript{Q\mu}{T}{\alpha}_{t_m}^{}\,\mathcal{M}_{\varkappa i j}^{\bs{J}^2} {\mathcal{H}}_\varkappa\nonumber
\end{align}
with $\prescript{dd}{}{\alpha}_{s}\equiv \prescript{dd}{}{\alpha}_{s}^{0}$. \\
In Appendix \ref{sec: Energy shift from static electromagnetic fields}, we derive that the induced energy shift of the system can be expressed through the same polarizabilities:
\begin{align}
 \Delta E\simeq &-\prescript{\mu}{}{\alpha}_{}\, \bs{\mathcal{J}}\cdot\bs{\mathcal{\mathcal{H}}} -\tfrac{1}{2} \prescript{dd}{}{\alpha}_{s}\, {\mathcal{E}}^2 -\tfrac{1}{2} \prescript{\mu\mu}{}{\alpha}_{s}\, {\mathcal{H}}^2 -\tfrac{1}{2} \prescript{dd}{}{\alpha}_{t}\, \mathcal{Q}^{\bs{J}^2}_{ij} {\mathcal{E}}_{i} {\mathcal{E}}_{j} -\tfrac{1}{2} \prescript{\mu \mu}{}{\alpha}_{t}\, \mathcal{Q}^{\bs{J}^2}_{ij} {\mathcal{H}}_{i} {\mathcal{H}}_{j}\nonumber\\
 & -\prescript{d}{PT}{\alpha}_{}^{}\, \bs{\mathcal{J}}\cdot\bs{\mathcal{E}} - \prescript{d\mu}{PT}{\alpha}_{s}^{}\, \bs{\mathcal{H}}\cdot\bs{\mathcal{E}} -\prescript{d\mu}{P}{\alpha}_{v}^{}\, (\bs{\mathcal{E}}\times\bs{\mathcal{H}})\cdot \bs{\mathcal{J}}- \prescript{d\mu}{PT}{\alpha}_{t}^{}\, \mathcal{Q}^{\bs{J}^2}_{ij}\mathcal{E}_i{\mathcal{H}}_j\label{eq: Delta E homo}
\end{align}
where the first line shows the terms that are even and the second line the terms that are odd under either $P$ or $T$.\\
We start by discussing the former. The first two terms are well-known and usually have the largest influence on the system. They describe the linear Zeeman- and the quadratic Stark effect. The third term is the quadratic Zeeman effect \cite{schiff1939theory}. It is best visible in diamagnetic systems where the linear Zeeman effect is absent. The fourth term is the Stark tensor-polarizability that can be relevant for systems with $J>1$ \cite{angel1968hyperfine}. The final term is an equivalent Zeeman tensor-interaction. It has been first considered in Ref. \cite{coulson1956quadrupole} as a way to induce an electric quadrupole moment through an applied magnetic field. \\

Now, we are going to discuss the symmetry-violating interactions, which are the main interest of this study.\\
The term $\prescript{d}{PT}{\alpha}_{}^{}\, \bs{\mathcal{J}}\cdot\bs{\mathcal{E}}$ is well known in the context of searches for the electric dipole moment of the electron (eEDM). It allows to restrict $P$- and $T$-violating physics by measuring the absence of linear stark shifts in paramagnetic atoms and molecules. This interaction does not uniquely arise from the eEDM but can also be caused by a variety of other $PT$-violating interactions, like Schiff moments, the scalar-pseudoscalar electron-nucleon coupling, or the electron interaction with the nuclear magnetic quadrupole moment. For a review on the topic, see Ref. \cite{chupp2019electric}. For purely diamagnetic systems, $\bs{\mathcal{J}}$ is zero. Such systems are, therefore, insensitive to the effects of $\prescript{d}{PT}{\alpha}_{}^{}$. \\

The remaining three terms in Eq. (\ref{eq: Delta E homo}) were first considered in Ref. \cite{feinberg1977parity}. The first of them, $\prescript{d\mu}{PT}{\alpha}_{s}^{}\, \bs{\mathcal{H}}\cdot\bs{\mathcal{E}}$, describes a direct coupling of the electric and magnetic fields similar to the magnetoelectric effect that can arise in some solids \cite{aroyo2013international}. In Ref. \cite{derevianko2010cp}, its effects were investigated in more detail. Through its scalar nature, $\prescript{d\mu}{PT}{\alpha}_{s}^{}$ contributes to para- and diamagnetic systems alike.\\
The properties of the term $(\bs{\mathcal{E}}\times\bs{\mathcal{H}})\cdot \bs{\mathcal{J}}$ were studied in Ref. \cite{frantsuzov1987vector}, and \cite{flambaum1992long}. Because of its antisymmetric structure, it is only nonvanishing when the angular momentum is misaligned with the direction of $\bs{\mathcal{H}}$ (and also the plane spanned by $\bs{\mathcal{E}}$ and $\bs{\mathcal{H}}$).\\
The final term $\prescript{d\mu}{PT}{\alpha}_{t}^{}\, \mathcal{Q}^{\bs{J}^2}_{ij}\mathcal{E}_i{\mathcal{H}}_j$ again describes a $PT$-odd electric magnetic field coupling. Apart from the initial proposal of this expression in Ref. \cite{feinberg1977parity}, we do not know of any other work discussing this coupling for static fields. This term requires a system with $J\geq 1$.\\

To measure these terms, one could directly measure the energy shift (\ref{eq: Delta E homo}) depending on the strength and direction of the applied fields. Alternatively, one could use a large sample and measure the induced polarization $\langle \bs{d}\rangle$, or magnetization $\langle \bs{\mu}\rangle$, from Eq. (\ref{eq: d mu homo}). An induced polarization in a solid can be measured with a precise voltage measurement as it was done in \cite{kim2015new} in the context of an eEDM search. Magnetizations of samples can, for example, be measured with a SQUID as it was done in another eEDM search in \cite{eckel2012limit}.\\

\subsubsection{Alignment of $\bs{\mathcal{J}}$ in externally applied fields}\label{sec: Alignment of J in externally applied fields}

The expectation value of the angular momentum $\bs{\mathcal{J}}=\br{n}\bs{J}\ke{n}$ is a vector whose orientation depends only on the unperturbed state $\ke{n}$ of the system. Even though $\bs{\mathcal{J}}$ is not directly dependent on the applied $\bs{\mathcal{E}}$ and $\bs{\mathcal{H}}$ fields, its direction can still be influenced by them.
This is because the energy shift in (\ref{eq: Delta E homo}) defines which $m$-substate in the $aJ$-manifold has the lowest energy. If the system possesses the ability to decay into its ground state $\ke{N_G}$, this will determine a particular $ \ke{aJm}$ state, which in turn implies a certain orientation of $\br{n}\bs{J}\ke{n}$.

For example, if we apply a strong magnetic field to a sample, the Zeeman coupling $\Delta E_{Z}=-\prescript{\mu}{}{\alpha}_{}\, \bs{\mathcal{J}}\cdot\bs{\mathcal{H}}$ dominates all the other terms. It is minimized if $\bs{\mathcal{J}}$ becomes maximally oriented to the direction of $\bs{\mathcal{H}}$. In this case, we can perform the replacement $\bs{\mathcal{J}}\rightarrow J\frac{\bs{\mathcal{H}}}{|\bs{\mathcal{H}}|}$. Here, $J$ is the usual angular momentum quantum number.\\
Alternatively, if magnetic fields are absent, then the application of an electric field will orient $\bs{\mathcal{J}}$ directly through the EDM coupling $\Delta E_{EDM}=-\prescript{d}{PT}{\alpha}_{}^{}\, \bs{\mathcal{J}}\cdot\bs{\mathcal{E}}$. We know from sec. \ref{sec: Full expressions of the induced moments} that an oriented angular momentum directly leads to an oriented magnetic moment $\langle \bs{\mu} \rangle$. It again could be picked up through magnetometry. This method of measuring the eEDM was first proposed in \cite{shapiro1968electric}.

\subsection{Static inhomogeneous fields}
\label{sec: Static inhomogeneous fields}
If we allow the electric field to vary in space ($\partial_i{\mathcal{E}}_j\neq0$), then the electric quadrupole terms will also contribute to the energy of the system. One possibility to apply strong field gradients to atoms or molecules is to embed them into a crystalline matrix \cite{satten1957effects, pryor1987artificial, kanorsky1998quadrupolar}.\\
In inhomogeneous fields, the dipole moments are extended by:
\begin{align*}
 \langle \bs{d} \rangle &= \dotsb+
 \prescript{dQ}{P}{\alpha}_{t_m}\,\mathcal{M}_{ij\varkappa}^{\bs{J}^2}\hat{\bs{e}}_i \partial_j {\mathcal{E}}_\varkappa+\left(\prescript{dQ}{PT}{\alpha}_{t_s}\,\mathcal{S}_{ij\varkappa}^{\bs{J}^3}+\prescript{dQ}{PT}{\alpha}_{v_m}\,\mathcal{M}_{ij\varkappa}^{\bs{J}}\right)\hat{\bs{e}}_i \partial_j {\mathcal{E}}_\varkappa\nonumber\\
 \langle \bs{\mu} \rangle &= 
 \dotsb +\prescript{\mu Q}{T}{\alpha}_{t_m}\,\mathcal{M}_{ij\varkappa}^{\bs{J}^2}\hat{\bs{e}}_i \partial_j {\mathcal{E}}_\varkappa+\left(\prescript{\mu Q}{}{\alpha}_{t_s}\,\mathcal{S}_{ij\varkappa}^{\bs{J}^3}+\prescript{\mu Q}{}{\alpha}_{v_m}\,\mathcal{M}_{ij\varkappa}^{\bs{J}}\right)\hat{\bs{e}}_i \partial_j {\mathcal{E}}_\varkappa
\end{align*}
The dots represent the terms from in Eq. (\ref{eq: d mu homo}). The inhomogeneous terms lead to an additional energy shift (see Appendix \ref{sec: Energy shift from static electromagnetic fields}):
\begin{align}
 \Delta E &= \dotsb -\prescript{Q}{}{\alpha}_{}^{}\,\mathcal{Q}^{\bs{J}^2}_{ij}\partial_i \mathcal{E}_j -\tfrac{1}{2}\prescript{QQ}{}{\alpha}_{ij\varkappa l}\,\partial_i \mathcal{E}_j\partial_\varkappa \mathcal{E}_l\label{eq: Delta E inhomo}\\
 &-\left(\prescript{\mu Q}{}{\alpha}_{t_s}\,\mathcal{S}_{ij\varkappa}^{\bs{J}^3}+\prescript{\mu Q}{}{\alpha}_{v_m}\,\mathcal{M}_{ij\varkappa}^{\bs{J}}\right)\mathcal{H}_i \partial_j {\mathcal{E}}_\varkappa\nonumber\\
 & -\prescript{dQ}{P}{\alpha}_{t_m}^{}\,\mathcal{M}_{ij\varkappa}^{\bs{J}^2}\mathcal{E}^{}_i \partial_j {\mathcal{E}}_\varkappa-\prescript{\mu Q}{T}{\alpha}_{t_m}^{}\,\mathcal{M}_{ij\varkappa}^{\bs{J}^2}\mathcal{H}^{}_i \partial_j {\mathcal{E}}_\varkappa\nonumber\\
 &-\left(\prescript{d Q}{PT}{\alpha}_{t_s}^{\ell}\,\mathcal{S}_{ij\varkappa}^{\bs{J}^3}+\prescript{d Q}{PT}{\alpha}_{v_m}^{}\,\mathcal{M}_{ij\varkappa}^{\bs{J}}\right)\mathcal{E}^{}_i \partial_j {\mathcal{E}}_\varkappa\nonumber
\end{align}
The expressions in the first two lines are classically allowed. The first two of these terms describe the interactions of static and induced electric quadrupole moments with the inhomogeneous electric field analogous to the linear and quadratic Stark and Zeeman effects. The next two terms describe the induction of an electric quadrupole moment through an applied magnetic field. That such an interaction, linear in $\bs{\mathcal{H}}$, exists was only considered quite recently in Ref. \cite{szmytkowski2012magnetic}.\\

The third and fourth lines in Eq. (\ref{eq: Delta E inhomo}) describes $P$, $T$, and $PT$-odd effects. We proceed by analyzing these terms in the scenario where the angular momentum structures $\mathcal{S}_{ij\varkappa}^{\bs{J}^3}$, $\mathcal{M}_{ij\varkappa}^{\bs{J}}$, and $\mathcal{M}_{ij\varkappa}^{\bs{J}^2}$ become aligned through the externally applied fields. The details about this process are illustrated in appendix \ref{sec:Alignment of MM in externally applied fields}.\\

The first of the two $ \mathcal{M}^{\bs{J}^2}_{ij\varkappa}$-dependent terms in Eq. (\ref{eq: Delta E inhomo}) describes an interaction that only violates $P$, but not $T$. It induces an energy shift of (see appendix \ref{sec:Alignment of MM in externally applied fields}):
\begin{align}
 \Delta E\propto \prescript{dQ}{P}{\alpha}_{t_m}^{}\, q\,(\bs{\mathcal{H}}\cdot \hat{\bs{n}})(\bs{\mathcal{E}}\times\bs{\mathcal{H}})\cdot \hat{\bs{n}}+\dotsb\label{eq: quadrupole P odd}
\end{align}
Here, $\hat{\bs{n}}$ is the direction, and $q$ is the strength of the quadrupole component of the inhomogeneous electrostatic potential $\phi^Q=q[(\bs{r}\cdot\hat{\bs{n}})^2-\tfrac{1}{3}\bs{r}^2]$. The proportionality factor for the expression above depends on the strength of the applied fields.\\
That a $P$-odd energy shift linear in $\bs{\mathcal{E}}$ could exist in an inhomogeneous electric field was first proposed in Ref. \cite{flambaum1992long}. Expressions of a similar form as Eq. (\ref{eq: quadrupole P odd}) have also been suggested in \cite{bouchiat2001atomic} and \cite{mukhamedjanov2005manifestations}. There, it was shown that the energy shift induced by the $P$-odd nuclear anapole moment is in a range that could realistically be measured with current technology.\\
Analogously to the derivation of $\Delta E$, the same effect also leads directly to an induced electric dipole moment:
\begin{align*}
 \langle \bs{d} \rangle \propto \prescript{dQ}{P}{\alpha}_{t_m}^{}\, q\,(\bs{\mathcal{H}}\cdot \hat{\bs{n}})(\bs{\mathcal{H}}\times \hat{\bs{n}})+\dotsb
\end{align*}
In contrast to the $PT$-odd electric dipole that results from$\prescript{d\mu}{PT}{\alpha}_{s}^{}$, this $P$-odd $T$-even one is oriented orthogonal to the direction of the applied magnetic field and stays invariant under $\bs{\mathcal{H}}\rightarrow -\bs{\mathcal{H}}$. \\

For the second $ \mathcal{M}^{\bs{J}^2}_{ij\varkappa}$-dependent term in Eq. (\ref{eq: Delta E inhomo}), we find:
\begin{align}
 \Delta E &\propto\prescript{\mu Q}{T}{\alpha}_{t_m}^{}\,b_2\,q\,(\bs{\mathcal{E}}\cdot \hat{\bs{n}})(\bs{\mathcal{H}}\times\bs{\mathcal{E}})\cdot \hat{\bs{n}}+\dotsb\label{eq: static Tope}\\
 \text{and }\langle \bs{\mu} \rangle &\propto\prescript{ \mu Q}{T}{\alpha}_{t_m}^{}\,b_2\,q\,(\bs{\mathcal{E}}\cdot \hat{\bs{n}})(\bs{\mathcal{E}}\times \hat{\bs{n}}) +\dotsb\nonumber
\end{align}
Unlike all effects discussed so far, this one is $T$-odd, but $P$-even (ToPe). To our knowledge, all atomic ToPe observables that were previously considered in the literature depend on dynamic processes (A review on ToPe physics can be found in Refs. \cite{hopkinson2002interferometric, khriplovich2012cp}). This would make this the first static mechanism for detecting such a symmetry violation. It could be possible to measure $\langle \bs{\mu}\rangle$ in a scheme similar to the one used in \cite{eckel2012limit}. \\

The last line of Eq. (\ref{eq: Delta E inhomo}) further contains two $PT$-odd terms. Here, we focus on the second one that depends on $\mathcal{M}_{ij\varkappa}^{\bs{J}}$. From it, we obtain the following observables:
\begin{align*}
 \Delta E &\propto q\,\prescript{d Q}{PT}{\alpha}_{v_m}^{}\,\left[(\bs{\mathcal{E}}\cdot \bs{\mathcal{H}})- (\bs{\mathcal{E}}\cdot \hat{\bs{n}})(\bs{\mathcal{H}}\cdot \hat{\bs{n}})\right]+\dotsb\\
  \langle \bs{d} \rangle &\propto q\,\prescript{d Q}{PT}{\alpha}_{v_m}^{}\, \left[\bs{\mathcal{H}}- \hat{\bs{n}}(\bs{\mathcal{H}}\cdot \hat{\bs{n}})\right]+\dotsb
\end{align*}
In contrast to the P-odd, T-even, and the P-even, T-odd energy shifts in Eqs. (\ref{eq: quadrupole P odd}), and
(\ref{eq: static Tope}), this expression is both odd under the inversion of $\bs{\mathcal{E}}$ and $\bs{\mathcal{H}}$. If $\bs{\mathcal{E}}$ and $\bs{\mathcal{H}}$ are parallel, but orthogonal to $\hat{\bs{n}}$, then the contribution from this interaction becomes very similar to $\prescript{d}{PT}{\alpha}_{}^{}$ and $\prescript{d\mu}{PT}{\alpha}_{s}^{}$. To our knowledge, this $PT$-odd term has not been discussed before.\\

We can conclude that a static quadrupolar electric field, as it can be present in crystals, allows one to use a combination of electric and magnetic fields to search for all three types of symmetry violations. Alternatively, to align the different angular momentum structures through applied fields, one could, of course, also prepare the system in a different state by optical pumping.

\subsection{Time-varying fields}
\label{sec: Time-varying fields}

If we allow for time-dependent $\bs{\mathcal{E}}$- and $\bs{\mathcal{H}}$-fields, all the terms in Eqs. (\ref{eq: <d>}), (\ref{eq: <mu>}), and Eq. (\ref{eq: <Q>}) contribute.
To measure these time-dependent observables, one could again directly pick up the induced magnetization $\langle \bs{\mu} \rangle$ or polarization $\langle \bs{d} \rangle$. \\
For example, $\langle\bs{\mu}\rangle=\prescript{\mu d}{P}{\alpha}_{v}^\ell\, ({\bs{\mathcal{E}}}^\ell\times\bs{\mathcal{J}})+\prescript{\mu d}{PT}{\alpha'}_{v}^\ell\, (\dot{\bs{\mathcal{E}}}^\ell\times\bs{\mathcal{J}})+\dotsb$ implies that applying an oscillating electric field to a polarized sample will induce an orthogonal magnetization. Its $P$-odd part oscillates in phase with the applied field, but its $PT$-odd part is out of phase. 
Another way to measure the time-dependent polarizabilities is to prepare the system in a well-defined state and observe the time evolution of its population when it is exposed to time-varying fields \cite{altuntacs2018demonstration}. \\

Finally, the effects of the $\langle \bs{d} \rangle$ and $\langle \bs{\mu} \rangle$ induced by electromagnetic waves propagating through a medium can be studied. 

 \begin{figure}
 \centering
 \includegraphics[width=0.55\textwidth]{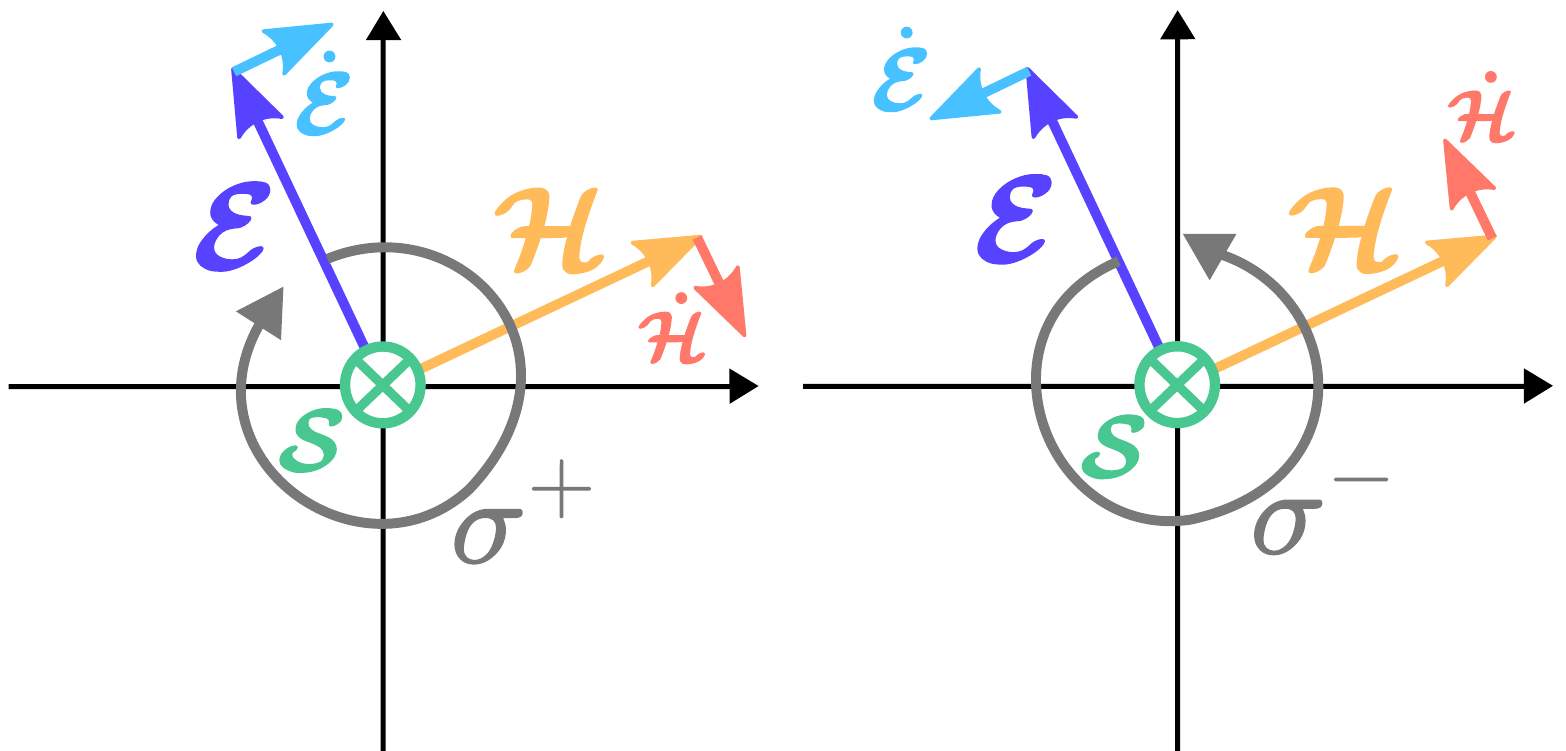}
 \caption{{\small Electric and magnetic fields of left ($\sigma^+$) and right ($\sigma^-$) circular polarized light. $\bs{\mathcal{S}}$ is the Poynting vector, defining the propagation direction of the light wave.}}
 \label{fig:polarization}
 \end{figure}

\noindent Let us consider the terms $ \langle \bs{d} \rangle = \prescript{dd}{}{\alpha}_{s}^{\ell}\, \bs{\mathcal{E}}^\ell+\prescript{d\mu}{P}{\alpha'}_{s}^{\ell}\, \dot{\bs{\mathcal{H}}}^\ell$. As shown in Fig. \ref{fig:polarization}, for a left circular polarized wave, $\dot{\bs{\mathcal{H}}}$ is parallel to $\bs{\mathcal{E}}$, while for a right circular polarized wave, they are antiparallel. From this, it follows that we can rewrite $\langle \bs{d}\rangle$ for the two polarizations $\sigma^\pm$ as $ \langle \bs{d} \rangle_\pm = \left(\prescript{dd}{}{\alpha}_{s}^{\ell}\, \mp \omega^\ell\,\prescript{d\mu}{P}{\alpha'}_{s}^{\ell}\, \right)\bs{\mathcal{E}}^\ell$. For dilute gases, the refractive index $n$ relates to the polarizability $\alpha$ through $n^2=\alpha+1$. This implies that the refractive index of left and right polarized light differ by an amount that is directly proportional to the size of the parity-violating potential $V^P$. This difference leads to a small optical rotation of light propagating through an atomic or (nonchiral) molecular gas.
The effect of $P$-odd optical rotation was successfully utilized to measure atomic parity violations in Bi, Pb, and Tl (\cite{bouchiat1997parity} and references within).
This is just one example of $P$, $T$ violation-induced birefringence and optical activity. In Refs. \cite{moskalev1986some, baryshevsky1994p} some others are discussed.\\

However, there is a caveat when dealing with time-dependent fields, which we discuss in the following section.

\section{Effects from the natural linewidth} \label{sec: Effects from the natural linewidth}

So far, we have neglected the effects from the natural linewidth of the excited states. These are not naturally contained in a semiclassical model for the interaction with the electromagnetic field. However, it is possible to model them by adding a non-Hermitian term to the Hamiltonian $H^0$ \cite{sakurai1967advanced}.

Under these considerations, the time-dependent perturbation expression in Eq. (\ref{eq:NdtM}) becomes \cite{vexiau2017dynamic}:
\begin{align}
\br{N}d_i\ke{N}(t)\simeq \br{n}&d_i\ke{n}\label{eq:NdtM Gamma}\\
- \sum_{l\neq 0}\sum_{k}&\left[\frac{\br{n}d_i\ke{k} \br{k}F^\ell\ke{n}}{\omega_{k n}-\tfrac{i}{2}\Gamma_k-\omega^\ell}+\frac{\br{n}F^\ell\ke{k}\br{k}d_i\ke{n} }{\omega_{k n}+\tfrac{i}{2}\Gamma_k+\omega^\ell}\right] e^{-i \omega^\ell t}\nonumber\\
+&\left[\frac{\br{n}d_i\ke{k} \br{k}(F^\ell)^\dagger\ke{n}}{\omega_{k n}-\tfrac{i}{2}\Gamma_k+\omega^\ell}+\frac{\br{n}(F^\ell)^\dagger\ke{k}\br{k}d_i\ke{n} }{\omega_{k n}+\tfrac{i}{2}\Gamma_k-\omega^\ell}\right] e^{i \omega^\ell t} \nonumber
\end{align}
where $\Gamma_k$ is the natural linewidth of the state $\ke{k}$.\\
Non-Hermitian Hamiltonians describe dissipative systems. The dissipation leads to an increase in entropy, which in turn defines an arrow of time. This effectively breaks $T$ symmetry without the need for a fundamental symmetry violation. This makes the unambiguous interpretation of $T$-odd observables in such a system difficult.

\subsection{Consequences for the induced electric dipole moment}
We will now illustrate how this effect manifests in the induced moments that we derived in the paper.\\

After evaluating Eq. (\ref{eq:NdtM Gamma}) analogously to what we did in sec. \ref{sec: Time dependent perturbation}, we get for $\br{N} d_i \ke{N}$:
\begin{align*}
\br{n}d_i\ke{n}-2Re\sum_{\ell \neq 0}\sum_k&\frac{\br{n}d_i\ke{k}\br{k}\big[\omega_{kn}V^{\ell}+i\dot{V}^{\ell}\big]\ke{n}}{(\omega^\ell)^2\Gamma_k^2+\left(\omega_{k n}^2-\tfrac{1}{4}\Gamma_k^2-(\omega^\ell)^2\right)^2}\left[i\omega^\ell\Gamma_k+\omega_{k n}^2-\tfrac{1}{4}\Gamma_k^2-(\omega^\ell)^2\right]\\
\end{align*}
This leads to the following expression for the induced dipole moment:
\begin{align*}
\br{N}d_i\ke{N}(t)\simeq& \br{n}d_i\ke{n}
\\
+ \sum_{l} \sum_k&\Big( Re \big[\prescript{dd}{}{{\alpha}}_{ij}^{\ell k}\big]-G^{\ell k}\, Im\big[\prescript{dd}{}{{\alpha}}_{ij}^{\ell k}\big]\Big)\mathcal{E}_{j}^\ell \nonumber\\
+ \sum_{l} \sum_k&\Big( Re \big[\prescript{d\mu}{}{{\alpha}}_{ij}^{\ell k}\big]-G^{\ell k}\, Im\big[\prescript{d\mu}{}{{\alpha}}_{ij}^{\ell k}\big]\Big)\mathcal{H}_{j}^\ell \nonumber\\
+ \sum_{l} \sum_k&\Big( Re \big[\prescript{dQ}{}{{\alpha}}_{ij\varkappa}^{\ell k}\big]-G^{\ell k}\, Im\big[\prescript{dQ}{}{{\alpha}}_{ij\varkappa}^{\ell k}\big]\Big)\partial_j\mathcal{E}_{\varkappa}^\ell \nonumber\\
+ \sum_{l} \sum_k&\Big( Im \big[\prescript{dd}{}{{\alpha'}}_{ij}^{\ell k}\big]+G^{\ell k}\, Re\big[\prescript{dd}{}{{\alpha'}}_{ij}^{\ell k}\big]\Big)\dot{\mathcal{E}}_{j}^\ell \nonumber\\
+\sum_{l} \sum_k&\Big( Im \big[\prescript{d\mu}{}{{\alpha'}}_{ij}^{\ell k}\big]+G^{\ell k}\, Re\big[\prescript{d\mu}{}{{\alpha'}}_{ij}^{\ell k}\big]\Big)\dot{\mathcal{H}}_{j}^\ell \nonumber\\
+\sum_{l} \sum_k&\Big( Im \big[\prescript{dQ}{}{{\alpha'}}_{ij\varkappa}^{\ell k}\big]+G^{\ell k}\, Re\big[\prescript{dQ}{}{{\alpha'}}_{ij\varkappa}^{\ell k}\big]\Big)\partial_j\dot{\mathcal{E}}_{\varkappa}^\ell 
\end{align*}
Here, we defined more general expressions for the polarizabilities that are valid for non-zero natural linewidths:
\begin{align*}
 \prescript{d \mu}{}{{\alpha}}_{ij}^{\ell k}&=2\frac{\left(\omega_{k n}^2-\tfrac{1}{4}\Gamma_k^2-(\omega^\ell)^2\right)\omega_{kn}}{(\omega^\ell)^2\Gamma_k^2+\left(\omega_{k n}^2-\tfrac{1}{4}\Gamma_k^2-(\omega^\ell)^2\right)^2}\br{n}d_i\ke{k}\br{k}\mu_j\ke{n}\\
 \prescript{d \mu}{}{{\alpha'}}_{ij}^{\ell k}&=-2\frac{\omega_{k n}^2-\tfrac{1}{4}\Gamma_k^2-(\omega^\ell)^2}{(\omega^\ell)^2\Gamma_k^2+\left(\omega_{k n}^2-\tfrac{1}{4}\Gamma_k^2-(\omega^\ell)^2\right)^2}\br{n}d_i\ke{k}\br{k}\mu_j\ke{n}\\
 G^{\ell k}&=\frac{\omega^\ell \Gamma_k}{\omega_{k n}^2-\tfrac{1}{4}\Gamma_k^2-(\omega^\ell)^2}
\end{align*}
If we take the static limit $\ell=0$, we get:
\begin{align*}
\sum_k\prescript{d \mu}{}{{\alpha}}_{ij}^{0 k}&=2\sum_k\frac{\br{n}d_i\ke{k}\br{k}\mu_j\ke{n}}{\omega_{k n}^2-\tfrac{1}{4}\Gamma_k^2}\omega_{kn}\quad\text{and}\quad G^{0 k}=0
\end{align*}
Therefore, in the case of static fields, everything considered in previous sections still holds. The only difference is that the size of all 1st order terms increases (slightly) through the difference in the denominator.\\

\subsection{Implication for time-dependent fields}

Returning to the general case $\omega_\ell\neq 0$, both the imaginary and real parts of the polarizabilities couple to the fields and their derivatives. We saw in previous sections that the observables for $T$-even and $T$-odd terms are different in that if one of them contains a field, the other necessarily contains its derivative. Now, with the inclusion of the natural linewidth, this is no longer true. Measuring a certain electromagnetic field response of a system no longer allows us to uniquely distinguish if it was caused by a $P$-odd $T$-even or by a $P$-odd $T$-odd effect because the same field configuration couples to both $_{P}{{\alpha}}_{}^\ell$, and $_{PT}{{\alpha}}_{}^\ell$. In the known standard model interactions, the strength of $P$-violating processes far exceeds that of $P$- and $T$-violating ones \cite{khriplovich1991parity}. Therefore, it is natural to assume the hierarchy $V^{P}\gg V^{PT}$. $P$-violating interactions can, therefore, pose a large background for experimental detections of $PT$-violating ones. To avoid this, we need the $G^{\ell k}$ factors to become small. This happens when $\omega^\ell$ is far detuned from all the resonances that contribute to the polarizability. As we have seen before, this can be achieved by choosing a low frequency. This has the potential disadvantage that also the field derivatives $\dot{\bs{\mathcal{H}}}^\ell$, and $\dot{\bs{\mathcal{E}}}^\ell$ become small in this limit. In the other case of very high frequencies, the polarizabilities $\prescript{}{}{{\alpha}}_{}^{\ell k}\sim \left(\omega^\ell\right)^{-2}$ themselves decrease rapidly.\\

In Refs. \cite{dzuba2018screening, tan2019screening}, it was shown that oscillating electric fields in neutral atoms and molecules are not well shielded at the nucleus. It was proposed to use this effect to measure nuclear EDMs. \\
In \cite{verma2020electron}, a different interesting new scheme was proposed to measure $P$, $T$ violations with clock transitions. It as well relies on utilizing an electric field oscillating close to the transition frequency. \\
The here presented effect of $P$- violation "mimicking" as $PT$-violation through the effects of the natural linewidths could pose an important background for these kinds of measurements.\\

The situation is even worse for $P$-even $T$-odd observables. Through the same effect, these mix with classical $P$-even, $T$-even potentials, whose size will be many orders of magnitude above the expected size of $T$-odd physics. It is, therefore, advantageous to use static polarizabilities such as $\prescript{\mu Q}{T}{\alpha}_{t_m}^{}$ to study ToPe effects.

\section{Conclusion and outlook}
In this work, we investigated how the violation of the symmetries of parity $P$ and time reversal $T$ leads to atomic and molecular responses to externally applied electric and magnetic fields that are classically forbidden. This led us to general expressions for the induced electric dipole (\ref{eq: <d>}), magnetic dipole (\ref{eq: <mu>}) and electric quadrupole moment (\ref{eq: <Q>}). We find that atoms and molecules can obtain electric dipole moments, not only through $PT$-violation (see appendix \ref{sec: T-violation}), but also through $P$ without $T$-violation in time-dependent (sec. \ref{sec: Induced V^P V^PT V^T}, \ref{sec: Effects from the natural linewidth}, appendix \ref{sec:electric dipole- magnetic dipole}), and even purely static fields (sec. \ref{sec: Static inhomogeneous fields}). Additionally, we derived Eqs. (\ref{eq: Delta E homo}), (\ref{eq: Delta E inhomo}) for the energy shift in static fields and Eq. (\ref{eq: <V>}) for the effects in time-varying ones. We connected most of these terms to effects that have previously been considered or used in experimental searches for atomic $P$- and $T$- violations (sec. \ref{sec: Static homogeneous fields}, \ref{sec: Static inhomogeneous fields}, \ref{sec: Time-varying fields}). Finally, we argued how the effects of natural linewidths could cause problems in searches of $T$- and $PT$- odd effects (sec. \ref{sec: Effects from the natural linewidth}).\\

To assess the feasibility of detecting the various terms presented in this work, it is, of course, necessary to evaluate their size as well as the signal-to-noise ratio that can be reached in a given experiment. However, making general statements about them is difficult as they depend heavily on the specific system under consideration. This is the level structure of the atom or molecule, which governs the size of $\omega_{nk}$, as well as the specific symmetry-violating effects that one wants to measure. Usually, one considers potentials $V^P$, $V^{PT}$, and $V^{T}$ that involve electron-nucleon contact interactions and, therefore, scale steeply with the number of nucleons and the overlap of the electrons with the nucleus \cite{khriplovich2012cp, bouchiat1997parity}. Therefore, searches involving very heavy atoms or molecules are usually the best choice \cite{arrowsmith2023opportunities}. However, there also exist models introducing long-range interactions $V^P$, $V^{PT}$, and $V^{T}$ through new light bosons that may be easier to restrict in lighter systems \cite{cong2024spin, khriplovich2012cp, karshenboim2010precision, jones2020probing, stadnik2018improved}. \\
To maximize the size of the polarizabilities, one also wants large $\bs{d}$, $\bs{\mu}$, and $\bs{Q}$ transition elements between close-lying states. Such configurations are usually easier to achieve in molecules than in atoms.\\After identifying promising systems using all these criteria, it will be necessary to perform atomic structure calculations in the future to evaluate the feasibility of measuring the various observables.\\

We hope that the treatment we presented in this paper can also provide a starting point for further investigations of other symmetry-violating effects that are not captured in our model due to its initial assumptions. In this paper, we only studied diagonal elements $\br{n}\bs{d}\ke{n}$. But with some modifications, one could also study the $P$ or $T$ forbidden dipole transitions $\br{n}\bs{d}\ke{m}$. It could also be interesting to extend the multipole expansion to include the magnetic quadrupole \cite{sushkov1984possibility} and toroidal moments \cite{sandars1993p}. One could also include a different class of new physics interactions by introducing $P$, $T$ odd potentials that are time-dependent or that couple directly to external fields. Finally, it would also be interesting to understand what happens if the field strength becomes too strong to justify the non-degenerate perturbative treatment that we used. 

\section{Acknowledgments}
This research was financed in whole or in part by
Agence Nationale de la Recherche (ANR) under the project ANR-21-CE30 -0028-01. \\
 A CC-BY public copyright license has been applied by the authors to the present document and will be applied to all subsequent versions up to the Author Accepted Manuscript arising from this submission, in accordance with the grant's open access conditions.\\
We thank O. Dulieu for fruitful discussions.

\appendix
\section{Interactions with electromagnetic fields }\label{sec: interaction with electromagnetic fields}
In the following, we define the electric and magnet fields $\bs{\mathcal{E}}$, and $\bs{\mathcal{H}}$ in respect to the electromagnetic vector and scalar potentials $\phi$ and $\bs{\mathcal{A}}$ in a way that allows us to separate the static and time-dependent components of the fields.
\begin{align}
 V^{EM}= -\bs{d}\cdot\bs{\mathcal{E}} - \bs{\mu}\cdot\bs{\mathcal{H}} - Q_{ij}\partial_i\mathcal{E}_j
\end{align}

The electric and magnetic fields $\bs{\mathcal{E}}(\bs{r},t) $ and $\bs{\mathcal{H}}(\bs{r},t)$ are defined through the electromagnetic scalar and vector potentials:
\begin{align*}
 \bs{\mathcal{E}}=-\nabla \phi - \partial_t \bs{\mathcal{A}}\,,\;\;\;\bs{\mathcal{H}}= \nabla\times \bs{\mathcal{A}}
\end{align*}
For a simple separation between time-independent and time-dependent terms, we use the gauge freedom to set $\partial_t\phi(\bs{r},t)=0$.\\
We define $\bs{\mathcal{A}}(\bs{r},t)$ in a rather general way, as a superposition of a static component and plane waves with arbitrary frequencies and polarizations:
\begin{align*}
 \bs{\mathcal{A}}=\bs{\mathcal{A}}^0 + \sum_{\ell\neq 0} Re\left(\bs{\mathcal{A}}^\ell e^{i\bs{k}^\ell\cdot \bs{r}} e^{-i\omega^\ell t}\right)
\end{align*}
$\bs{\mathcal{A}}^0$ is the (real) static component of the vector potential ($\partial_t \bs{\mathcal{A}}^0=0$). $\bs{k}^\ell$ is the wavevector and $\omega^\ell$ is the angular frequency of the wave. The real part $Re$ is defined as $2 Re\,A=A+A^*$, with $A^*$ being the complex conjugate to $A$. Throughout this paper, both upper and lower indices are used for ease of notation without the positioning carrying a specific meaning. From these choices, the electric field is given by:
\begin{align}
 \bs{\mathcal{E}}&=\bs{\mathcal{E}}^0 + \sum_{\ell\neq 0}\bs{\mathcal{E}}^\ell(t)\nonumber\\
 &=-\nabla \phi + \sum_{\ell\neq 0}Re\left(i\omega^\ell\bs{\mathcal{A}}^\ell e^{i\bs{k}^\ell\cdot \bs{r}} e^{-i\omega^\ell t}\right)\label{eq: E E0 Et}
 \end{align}
with the first term describing the static and the second the time-dependent field component. Accordingly, for the magnetic field $\bs{\mathcal{H}}$:
\begin{align}
 \bs{\mathcal{H}}&=\bs{\mathcal{H}}^0 + \sum_{\ell\neq 0}\bs{\mathcal{H}}^\ell(t)\nonumber\\
 &=\nabla \times \bs{\mathcal{A}}^0 +\sum_{\ell\neq 0}Re\left(i\left(\bs{k}^\ell\times\bs{\mathcal{A}}^\ell\right) e^{i\bs{k}^\ell\cdot \bs{r}} e^{-i\omega^\ell t}\right)\label{eq: H H0 Ht}
\end{align}
The Jacobian of $\bs{\mathcal{E}}$ is given by:
\begin{align}
\partial_i\mathcal{E}_j&= - \partial_i\partial_j \phi - \sum_{\ell\neq 0}Re\left(\omega^\ell k_i^\ell \mathcal{A}_j^\ell e^{i\bs{k}^\ell\cdot \bs{r}} e^{-i\omega^\ell t}\right)
\label{eq: E H derivatives}
 \end{align}
We solve the system in the inertial center of mass frame of the atom (molecule). For a moving system, it is necessary to transform the field values and frequencies accordingly.\\\\
In the multipole approximation, only the field values at the center of mass contribute \cite{steck2007quantum}:
\begin{align*}V^{EM}=-\bs{d}\cdot\bs{\mathcal{E}}|_{\bs{r}=0}-\bs{\mu}\cdot\bs{\mathcal{H}}|_{\bs{r}=0}-Q_{ij}(\partial_i\mathcal{E}_j)|_{\bs{r}=0}
\end{align*}
For the time-dependent field components, this implies $e^{i\bs{k}^\ell\cdot \bs{r}} =1$. For simplicity of notation, we will not explicitly indicate this limit elsewhere. 

\section{The time-reversal operator $T$}\label{sec: The time-reversal operator T}
As an antiunitary operator, the time reversal operator $T$ has some unintuitive properties with respect to the usual bra-ket notation (see, for example, Ref. \cite{messiah2014quantum}). For linear Operators $L$, the bra-ket notation is associative in the sense that:
\begin{align*}
   \br{\varphi} (L \ke{\psi}) = ( \br{\varphi} L )\ke{\psi}\equiv \br{\varphi} L \ke{\psi} 
\end{align*}
For an antiunitary operator, this is not true. There we have:
\begin{align*}
   \br{\varphi} (A \ke{\psi}) = \left[ ( \br{\varphi} A )\ke{\psi} \right]^*
\end{align*}
For ease of notation, we do not indicate the parentheses explicitly in the rest of the paper. Instead, we define:
\begin{align*}
   \br{\varphi} A \ke{\psi} \equiv \br{\varphi} (A \ke{\psi})
\end{align*}
For the action of the inverse of the antiunitary time reversal operator $T^\dagger$ onto a $\br{\text{bra}}$ follows:
\begin{align*}
  \br{a J m}T^\dagger\ke{k}&\equiv\br{a J m}(T^\dagger\ke{k})=\big[ (\br{k}T)\ke{aJm}\big]^*=\br{k}(T\ke{aJm})\\
  &=(-1)^{J-m}\langle k |aJ -m\rangle=(-1)^{J-m}\big[\langle aJ-m| k\rangle \big]^*
\end{align*}

\section{Analyzing the polarizabilities in terms of parity and time reversal} \label{app: Analyzing the polarizabilities in terms of parity and time reversal}
In this appendix we analyze the generalized polarizabilities introduced in sec. \ref{sec: P and T odd polarizabilities} regarding their transformation behavior under $P$ and $T$. From this we will see that around three-quarters of all possible terms in Eq. (\ref{eq: <d> P PT T}) necessarily vanish in all systems that obey our initial assumptions made in sec. \ref{sec: Requirements on the system}.
\subsection{$P$ and $T$ properties of polarizabilities in zero field: 0th order in $V^{EM}$}
\label{sec: 0st order}
We start by discussing the vector polarizabilities $\prescript{d}{}{\alpha}_{i}^{}$, $\prescript{d}{P}{\alpha}_{i}^{}$, $\prescript{d}{PT}{\alpha}_{i}^{}$, and $\prescript{d}{T}{\alpha}_{i}^{}$. Because these arise in the 0th order perturbation in $V^{EM}$, they are independent of $\bs{\mathcal{E}}$, and $\bs{\mathcal{H}}$. As such, they describe a permanent electric dipole moment (EDM).

\subsubsection{P-violation}\label{sec: P-violation}
It can easily be shown that $\prescript{d}{}{\alpha}_{i}^{}= 0$ due to the odd parity of $\bs{d}$. Using the properties given in Table \ref{tab:my_table} we get:
\begin{align}
 \prescript{d}{}{\alpha}_{i}^{}= \br{n}d_i\ke{n}=\br{n}PP\,d_i\,PP\ke{n}=(\pm\br{n})(-d_i)(\pm\ke{n})=-\br{n}d_i\ke{n}\label{eq: d P-odd}
\end{align}
Thus $\prescript{d}{}{\alpha}_{i}^{} =0$.\\
To get a non-zero expectation value, we need to mix states of different parity. Effectively, this is what happens when we introduce the perturbation trough $V^{P}$ or $V^{PT}$. In this case:
\begin{align*}
 \prescript{d}{P}{\alpha}_{i}^{}&= 2 \sum_k \br{n} PP d_i PP\ke{k}\br{k}PP V^{P} PP\ke{n}\,\omega_{kn}^{-1}\\
 &=2 \sum_k \br{n} d_i \ke{k}\br{k} V^{P} \ke{n}\,\omega_{kn}^{-1}=\prescript{d}{P}{\alpha}_{i}^{}
\end{align*}
Therefore $\prescript{d}{P}{\alpha}_{i}^{}$ is not forbidden by parity arguments.

\subsubsection{T-violation} 
\label{sec: T-violation}
It is further possible to show that $\prescript{d}{}{\alpha}_{i}^{}$ also vanishes due to reasons of $T$-symmetry. This fact was first demonstrated by Lee and Yang \cite{lee1957elementary}. Through the Wigner Eckart theorem, we know that the vector operator $\bs{d}$ has to be oriented along the total angular momentum $\bs{J}$:
\begin{align}
 \br{a J m}d_i\ke{a J m}=c_J^a \br{a J m}J_i\ke{a J m}\label{eq: lee yang}
\end{align}
with a real $m$-independent constant $c_J^a$ \cite{steck2007quantum}. 
We can now introduce the time reversal operator $T$ to the LHS of Eq. (\ref{eq: lee yang}). Using properties given in Table \ref{tab:my_table} we get:
\begin{align*}
 &\br{a J m}d_i\ke{a J m}=\br{a J m}T^\dagger Td_iT^\dagger T\ke{a J m}\\
 =&\left(\br{a J -m} Td_iT^\dagger \ke{a J -m}\right)^* =\left(\br{a J -m} d_i \ke{a J -m}\right)^*
\end{align*}
where the complex conjugation originates from the antilinear properties of $T$ (see appendix \ref{sec: The time-reversal operator T}).
Analogously, we can show for the RHS of Eq. (\ref{eq: lee yang}):
\begin{align*}
 &\br{a J m}J_i\ke{a J m}=-\left(\br{a J -m} J_i \ke{a J -m}\right)^*
\end{align*}
where we used that $J_i$ is $T$-odd.
We can now rename $-m\rightarrow m$ and use the fact that both $d_i$, and $J_i$ are Hermitian, to arrive at:
\begin{align*}
 \br{a J m}d_i\ke{a J m}=-c_J^a \br{a J m}J_i\ke{a J m}
\end{align*}
Then the comparison with Eq. (\ref{eq: lee yang}) clearly implies $ \br{a J m}d_i\ke{a J m}=0$.\\

We now show how this argument fails under the introduction of a $T$-odd potential $V^T$ or $V^{PT}$:
\begin{align*}
 Re\left(\prescript{d}{PT}{\alpha}_{i}^{}\right)=&2Re\sum_k\br{a J m}d_i\ke{k}\br{k}V^{PT}\ke{a J m}\omega_{nk}^{-1}\\
 =&\br{a J m}\sum_k\left[\left(d_i\ke{k}\br{k}V^{PT}\right)+\left(V^{PT}\ke{k}\br{k}d_i\right)\right]\ke{a J m}\omega_{nk}^{-1}\\
 \equiv&\br{a J m}D_i\ke{a J m}\\
\end{align*}
Here we introduced the Hermitian vector operator $D_i$. This operator is odd under time reversal (and even under Parity):
\begin{align*}
 T D_i T^\dagger&=\sum_k\left[\left(Td_i T^\dagger T \ke{k}\br{k} T^\dagger T V^{PT} T^\dagger \right)+\left( T V^{PT} T^\dagger T \ke{k}\br{k} T^\dagger T d_i T^\dagger \right)\right]\omega_{nk}^{-1}\\
 &=-\sum_k\left[\left(d_i \ke{k}\br{k} V^{PT} \right)+\left( V^{PT} \ke{k}\br{k} d_i \right)\right]\omega_{nk}^{-1}\\
\end{align*}
where we used that $V^{PT}$ is $T$-odd while $d_i$ and $\sum_k \ke{k}\br{k}\omega_{nk}^{-1}$ are $T$-even.
Previously, for the Lee-Yang argument, the contradiction arose because $d$ and $J$, which are proportional to each other, transform differently under $T$. Now that both $D$ and $J$ are $T$-odd, the contradiction vanishes.\\
The combined requirements on parity and time reversal leave $\prescript{d}{PT}{\alpha}_{i}^{}$ as the only possible source for an EDM and tells us that such a property is directly proportional to the size of $V^{PT}$.

\subsubsection{Magnetic dipole moment $\bs{\mu}$}\label{sec: magnetic dipole moment mu}
Contrary to the electric dipole moment, the 0st-order expectation value of the magnetic dipole moment is already classically allowed. We can also easily see this from the fact that $D_i$ and $\mu_i$ transform in the same way under $P$ and $T$. From this and our derivation above, it directly follows that the value of the static magnetic dipole moment is insensitive to the effects of $V^{P}$, $V^{T}$, and $V^{PT}$. Therefore, $\prescript{\mu}{}{\alpha}_{i}^{}$ is the only non-zero contribution to the static magnetic dipole moment.
Through the Wigner-Eckart theorem, we can express it as 
\begin{align*}
\prescript{\mu}{}{\alpha}_{i}^{}=\prescript{\mu}{}{\alpha}_{}^{} \br{n}{J}_i\ke{n} = \prescript{\mu}{}{\alpha}_{}^{} \mathcal{J}_i 
\end{align*}
where $\prescript{\mu}{}{\alpha}_{}^{}$ is a real constant and equivalent to the $c_J^a$ factor that we introduced above. We further define the notation
\begin{align*}
  \bs{\mathcal{J}}\equiv\br{n}\bs{J}\ke{n}
\end{align*}
for the expectation value of the total angular momentum in a given state.
\subsubsection{Electric quadrupole moment $Q_{ij}$}
The electric quadrupole operator ${Q}_{ij}$ is both symmetric and traceless. The expectation value of such an object can only be proportional to the angular momentum quadrupole \cite{varshalovich1988quantum}: 
\begin{align}
\mathcal{Q}^{\bs{J}^2}_{ij}\equiv\br{n}\tfrac{1}{2}\Big[J_i J_j+J_j J_i-\tfrac{2}{3}\delta_{ij}\sum_lJ_l J_l\Big]\ke{n} \label{eq: def Q}
\end{align}
 Therefore, we can write:
\begin{align*}
\langle Q_{ij}\rangle=\prescript{Q}{}{\alpha}_{}^{}\,\mathcal{Q}^{\bs{J}^2}_{ij}
\end{align*}

\subsection{$P$ and $T$ properties of polarizabilities in 1st order in $V^{EM}$}\label{sec: 1st order in V^EM}
We now continue by analyzing the rank-2 tensor polarizabilities $\prescript{}{}{\alpha}_{ij}^{\ell }$ by using the techniques that we developed in the previous section. 
\subsubsection{Restrictions from parity}
As previously in Eq. (\ref{eq: d P-odd}), we can analyze the $\prescript{}{}{\alpha}_{ij}^{\ell }$ that are defined in Eq. (\ref{eq: alpha PT d mu}) by introducing $P$ into the expression. It can easily be seen that terms that contain an odd number of $P$-odd operators will vanish ($\prescript{d\mu}{T}{\alpha}_{ij}^{\ell }$ for example contains the $P$-even operators $\mu_j$, and $V^T$ and the $P$-odd operator $d_i$). Therefore, half of all possible polarizabilities vanish due to parity reasons.\\

\subsubsection{Tensor decomposition of $\prescript{}{}{\alpha}_{ij}^{\ell}$}
\label{sec: Tensor decomposition}

Before we are able to analyze $\prescript{}{}{\alpha}_{ij}^{\ell}$ in terms of time reversal, we first need to understand the structure of these (Eucledian) rank-2 tensors. Such objects can be decomposed into a sum of three components that transform as scalar, antisymmetric vector, and traceless symmetric tensors respectively \cite{varshalovich1988quantum}: 
\begin{align*}
 3\otimes 3 &= \quad\quad\underset{\text{scalar}}{1} &\oplus& \quad\quad\;\underset{\text{vector}}{3} &\oplus& \quad\quad\quad\quad\quad\quad\;\underset{\text{tensor}}{5} \nonumber\\
 \alpha_{ij}^{\ell } &= \tfrac{1}{3} \sum_\varkappa\alpha_{\varkappa \varkappa}^{\ell }\;\delta_{ij} &+& \tfrac{1}{2}\left[\alpha_{ij}^{\ell } - \alpha_{ji}^{\ell }\right] &+& \Big(\tfrac{1}{2}\left[\alpha_{ij}^{\ell } + \alpha_{ji}^{\ell }\right] - \tfrac{1}{3}\sum_\varkappa(\alpha^{\ell }_{\varkappa\varkappa})\delta_{ij}\Big)
\end{align*}
with $\delta_{ij}$ being the Kronecker delta. Angular momentum algebra tells us that the expectation value of a vector operator is necessarily oriented along $\bs{\mathcal{J}}$, and the one of a traceless rank-2 tensor needs to be proportional to $\mathcal{Q}^{\bs{J}^2}_{ij}$ \cite{varshalovich1988quantum}. This allows us to express the tensor ${\alpha}_{ij}^{\ell }$, with only three constants ${\bar{\alpha}}_s^{\ell }$ $,\prescript{}{}{\bar{\alpha}}_v^{\ell }$, and $\prescript{}{}{\bar{\alpha}}_t^{\ell }$:
\begin{align}
 \prescript{}{}{\alpha}_{ij}^{\ell }=\delta_{ij}\; \prescript{}{}{\bar{\alpha}}_s^{\ell }+\varepsilon_{ij\varkappa}\mathcal{J}_\varkappa\; \prescript{}{}{\bar{\alpha}}_v^{\ell }+\mathcal{Q}^{\bs{J}^2}_{ij}\; \prescript{}{}{\bar{\alpha}}_t^{\ell }\label{eq: tensor decomposition}
\end{align}
with $\varepsilon_{ijk}$ the Levi-Civita symbol. A non-zero value for the vector term requires the system to possess $J\geq \tfrac{1}{2}$, while the tensor term will only exist for systems with $J\geq 1$. ${\bar{\alpha}}_s^{\ell }$ $,\prescript{}{}{\bar{\alpha}}_v^{\ell }$, and $\prescript{}{}{\bar{\alpha}}_t^{\ell }$ depend on the quantum numbers $a$ and $J$, but not $m$. We use the bar above them to indicate that they are complex-valued. In the following, we will construct expressions for the tensor decomposition in terms of purely real constants.\\

\subsubsection{Restrictions from time reversal}\label{sec:electric dipole- magnetic dipole}
To further eliminate terms, we can again make use of their transformation under time reversal similar to \cite{zhizhimov1982p} and to what we did in sec. \ref{sec: T-violation}. \\
We will perform this procedure on the example of $\prescript{d\mu}{}{\alpha}_{ij}^{\ell}$ (we ignore for now that this polarizability vanishes for Parity reasons). Again, we can decompose $\prescript{d\mu}{}{\alpha}_{ij}^{\ell}$ into its irreducible components:
 \begin{align}
 \prescript{d\mu}{}{\alpha}_{ij}^{\ell}&=\delta_{ij}\; \prescript{d\mu}{}{\bar{\alpha}}_s^{\ell}+\varepsilon_{ij\varkappa}\br{a J m}J_\varkappa\ke{a J m}\; \prescript{d\mu}{}{\bar{\alpha}}_v^{\ell}\label{eq: dmu LHS=RHS}\\
&+\br{a J m}\tfrac{1}{2}\left[J_i J_j+J_j J_i-\tfrac{2}{3}\delta_{ij}\sum_lJ_lJ_l\right]\ke{a J m}\; \prescript{d\mu}{}{\bar{\alpha}}_t^{\ell}\nonumber
\end{align}
 We introduce $1=T^\dagger T$ into the LHS that we previously defined trough Eq. (\ref{eq: def alpha}) to get:
 \begin{align}
 \prescript{d\mu}{}{\alpha}_{ij}^{\ell}&=2\sum_k\frac{\br{a J m} T^\dagger T d_i T^\dagger T \ke{k}\br{k} T^\dagger T \mu_j T^\dagger T \ke{a J m}}{\omega_{kn}^2-(\omega^\ell)^2}\omega_{kn}\label{eq: dmu LHS}\\
 &=-2\sum_k\frac{\left(\br{a J -m} d_i\ke{k}\br{k}\mu_j\ke{a J -m}\right)^*}{\omega_{kn}^2-(\omega^\ell)^2}\omega_{kn} \nonumber
\end{align}
The RHS of Eq. (\ref{eq: dmu LHS=RHS}) transforms as:
\begin{align}
 \prescript{d\mu}{}{\alpha}_{ij}^{\ell} =& \delta_{ij}\; \prescript{d\mu}{}{\bar{\alpha}}_s^{\ell}-\varepsilon_{ij\varkappa}\br{a J -m}J_\varkappa\ke{a J -m}\; \prescript{d\mu}{}{\bar{\alpha}}_v^{\ell}\label{eq: dmu RHS}\\
&+\br{a J -m}\tfrac{1}{2}\left[J_i J_j+J_j J_i-\tfrac{2}{3}\delta_{ij}\sum_lJ_lJ_l\right]\ke{a J -m}\; \prescript{d\mu}{}{\bar{\alpha}}_t^{\ell}\nonumber
\end{align}
The important observation here is that the vector part gets a minus sign in relation to the scalar and tensor parts.\\
We can now equate Eq. (\ref{eq: dmu LHS}) with Eq. (\ref{eq: dmu RHS}) and rename $-m\rightarrow m$ to obtain:
\begin{align*}
   &-\Bigg[2\sum_k\frac{\br{a J m} d_i\ke{k}\br{k}\mu_j\ke{a J m}}{\omega_{kn}^2-(\omega^\ell)^2}\omega_{kn}\Bigg]^* \\
   &=\delta_{ij}\; \prescript{d\mu}{}{\bar{\alpha}}_s^{\ell}-\varepsilon_{ij\varkappa}\br{a J m}J_\varkappa\ke{a J m}\; \prescript{d\mu}{}{\bar{\alpha}}_v^{\ell}\\
&\quad\;+\br{a J m}\tfrac{1}{2}\left[J_i J_j+J_j J_i-\tfrac{2}{3}\delta_{ij}\sum_lJ_lJ_l\right]\ke{a J m}\; \prescript{d\mu}{}{\bar{\alpha}}_t^{\ell}\nonumber\\
 \Rightarrow&\left(\prescript{d\mu}{}{\alpha}_{ij}^{\ell}\right)^*=-\delta_{ij}\; \prescript{d\mu}{}{\bar{\alpha}}_s^{\ell}+\varepsilon_{ij\varkappa}\mathcal{J}_\varkappa\; \prescript{d\mu}{}{\bar{\alpha}}_v^{\ell}-\mathcal{Q}^{\bs{J}^2}_{ij}\; \prescript{d\mu}{}{\bar{\alpha}}_t^{\ell}
\end{align*}
The comparison with Eq. (\ref{eq: dmu LHS=RHS}) finally allows us to express $\prescript{d\mu}{}{\alpha}_{ij}^{\ell}$ in terms of real coefficients:
 \begin{align*}
 \prescript{d\mu}{}{\alpha}_{ij}^{\ell}=i\delta_{ij}\; \prescript{d\mu}{}{\alpha}_s^{\ell}+\varepsilon_{ij\varkappa}\mathcal{J}_\varkappa\; \prescript{d\mu}{}{\alpha}_v^{\ell}+i\mathcal{Q}^{\bs{J}^2}_{ij}\; \prescript{d\mu}{}{\alpha}_t^{\ell}
\end{align*}
with $\prescript{d\mu}{}{\alpha}_{v}^{\ell}\equiv Re\left(\prescript{d\mu}{}{\bar{\alpha}}_{v}^{\ell}\right)$. The decomposition of $\prescript{d\mu}{}{\alpha'}_{v}^{\ell}$, that only differs by a factor of $-\omega_{kn}$ follows in the exact same way.\\
If we perform an analog treatment on $\prescript{dd}{}{\alpha}_{ij}^{\ell}$, everything is basically the same except that we do not get the global minus sign in Eq. (\ref{eq: dmu LHS}) due to the different $T$-transformation behavior of $d_j$ and $\mu_j$. Therefore:
\begin{align}
 \left(\prescript{dd}{}{\alpha}_{ij}^{\ell}\right)^*&=\delta_{ij}\; \prescript{dd}{}{\bar{\alpha}}_s^{\ell}-\varepsilon_{ij\varkappa}\mathcal{J}_\varkappa\; \prescript{dd}{}{\bar{\alpha}}_v^{\ell}+\mathcal{Q}^{\bs{J}^2}_{ij}\; \prescript{dd}{}{\bar{\alpha}}_t^{\ell} \nonumber\\
 \Rightarrow\prescript{dd}{}{\alpha}_{ij}^{\ell}&=\delta_{ij}\; \prescript{dd}{}{\alpha}_s^{\ell}+i\varepsilon_{ij\varkappa}\mathcal{J}_\varkappa\; \prescript{dd}{}{\alpha}_v^{\ell}+\mathcal{Q}^{\bs{J}^2}_{ij}\; \prescript{dd}{}{\alpha}_t^{\ell}
 \label{eq:alphadd}
\end{align}
\subsubsection{Behavior of the $d\mu$-terms under time reversal}
\label{sec: Behavior of the d mu terms under time reversal}
We now demonstrate that the argument we presented above for the transformation of $\prescript{d\mu}{}{\alpha}_{ij}^{\ell}$ and $\prescript{dd}{}{\alpha}_{ij}^{\ell}$ under time reversal $T$ still applies for the more complex $P$, $T$ odd polarizability expressions like $\prescript{d\mu}{P}{\alpha}_{ij}^{\ell}$.\\
We again start by decomposing $\prescript{d\mu}{P}{\alpha}_{ij}^{\ell}$ into its irreducible components:
 \begin{align}
 \prescript{d\mu}{P}{\alpha}_{ij}^{\ell}&=\delta_{ij}\; \prescript{d\mu}{P}{\bar{\alpha}}_s^{\ell}+\varepsilon_{ij\varkappa}\br{a J m}J_\varkappa\ke{a J m}\; \prescript{d\mu}{P}{\bar{\alpha}}_v^{\ell}\label{eq: dmu VP LHS=RHS}\\
&+\br{a J m}\tfrac{1}{2}\left[J_i J_j+J_j J_i-\tfrac{2}{3}\delta_{ij}\sum_lJ_lJ_l\right]\ke{a J m}\; \prescript{d\mu}{P}{\bar{\alpha}}_t^{\ell}\nonumber
\end{align}
It is clear that the RHS will transform exactly the same as in Eq. (\ref{eq: dmu RHS}). For $\prescript{d\mu}{P}{\alpha}_{ij}^{\ell}$ on the LHS, we can introduce $1=T^\dagger T$ into Eq. (\ref{eq: alpha d mu}). After using the transformation behaviors of the different operators given in table \ref{tab:my_table}, we get:
\begin{align}
 \prescript{d\mu}{P}{\alpha}_{ji}^{\ell}=-2 \sum_{k,\,l}\Bigg[ &\frac{\br{a J -m}V^{P}\ke{l}\br{l}d_i\ke{k}\br{k}\mu_j\ke{a J -m}}{(\omega_{kn}^2-(\omega^\ell)^2)\omega_{ln}}\omega_{kn}\label{eq: alpha dmu TT}\\
+& \frac{\br{a J -m}d_i\ke{k}\br{k}\mu_j\ke{l}\br{l}V^{P}\ke{a J -m}}{(\omega_{kn}^2-(\omega^\ell)^2)\omega_{ln}}\omega_{kn}\nonumber\\
  +&\frac{\br{a J -m}d_i\ke{k}\br{k}V^{P}\ke{l}\br{l}\mu_j\ke{a J -m}}{(\omega_{kn}^2-(\omega^\ell)^2)(\omega_{ln}^2-(\omega^\ell)^2)}\left(\omega_{kn}\omega_{ln}+(\omega^\ell)^2\right)\Bigg]^*\nonumber
\end{align}
Equating this with the transformed version of the RHS of Eq. (\ref{eq: dmu VP LHS=RHS}) and renaming $-m\rightarrow m$, we get:
\begin{align*}
 \left(\prescript{d\mu}{P}{\alpha}_{ij}^{\ell}\right)^*=-\delta_{ij}\; \prescript{d\mu}{P}{\bar{\alpha}}_s^{\ell}+\varepsilon_{ij\varkappa}\mathcal{J}_\varkappa\; \prescript{d\mu}{P}{\bar{\alpha}}_v^{\ell}-\mathcal{Q}^{\bs{J}^2}_{ij}\; \prescript{d\mu}{P}{\bar{\alpha}}_t^{\ell}
\end{align*}
This tells us that the symmetric part is purely imaginary while the antisymmetric vector component is real. \\
It is now also clear that when considering $\prescript{d\mu}{PT}{\alpha}_{ij}^{\ell }$ instead of $\prescript{d\mu}{P}{\alpha}_{ji}^{\ell }$, the RHS of Eq. (\ref{eq: alpha dmu TT}) will change by a global minus sign due to the fact that $V^{PT}$ is $T$-odd. That way, we have:
\begin{align}
 \left(\prescript{d\mu}{PT}{\alpha}_{ij}^{\ell }\right)^*=\delta_{ij}\; \prescript{d\mu}{PT}{\bar{\alpha}}_s^{\ell }-\varepsilon_{ij\varkappa}\mathcal{J}_\varkappa\; \prescript{d\mu}{PT}{\bar{\alpha}}_v^{\ell }+\mathcal{Q}^{\bs{J}^2}_{ij}\; \prescript{d\mu}{PT}{\bar{\alpha}}_t^{\ell }\label{eq: d mu PT tensor decomp}
\end{align}
\\
We conclude that polarizabilities containing an even number of $T$-odd operators possess purely real scalar and tensor components and a purely imaginary vector component. Polarizabilities with an odd number of $T$-odd operators, on the other hand, have a real vector component but imaginary scalar and tensor parts. From the definitions in \ref{sec: P and T odd polarizabilities}, we can see that the polarizabilities $\alpha$ that couple to the fields need to be necessarily real while the polarizabilities $\alpha'$ that couple to the field derivatives need to be imaginary.

\subsubsection{Dipole-quadrupole terms: rank-3 tensor decomposition}
\label{sec: Rank-3 tensor decomposition}
We continue by examining the rank-3 polarizability tensors. These arise in Eqs. (\ref{eq:NdN}), and (\ref{eq:NmuN}) as cross polarizabilities between electric and magnetic dipoles and the electric quadrupole moment.\\

As a symmetric tensor, the quadrupole operator can only have 6 independent elements, leading to an $\alpha_{ij\varkappa}^{\ell}$ with $3\otimes 6=18$ elements. The decomposition of a rank-3 tensor that is symmetric in its last two components under $SU(2)$ was derived in Ref. \cite{itin2022decomposition}:
\begin{align*}
 18=\underset{vector}{3}\oplus \underset{\substack{tensor\\(octupole)}}{7}\oplus \underset{vector}{3}\oplus \underset{\substack{tensor\\(quadrupole)}}{5}
\end{align*}
where the first two tensors are symmetric and the second two are of mixed symmetry. This allows us to write the rank-3 polarizabilities as:
\begin{align*}
\prescript{}{}{\alpha}_{ij\varkappa}^{\ell}=\mathcal{S}_{ij\varkappa}^{\bs{J}}\,\prescript{}{}{\bar{\alpha}}_{v_s}^{\ell}+\mathcal{S}_{ij\varkappa}^{\bs{J}^3}\,\prescript{}{}{\bar{\alpha}}_{t_s}^{\ell}+\mathcal{M}_{ij\varkappa}^{\bs{J}}\,\prescript{}{}{\bar{\alpha}}_{v_m}^{\ell}+\mathcal{M}_{ij\varkappa}^{\bs{J}^2}\,\prescript{}{}{\bar{\alpha}}_{t_m}^{\ell}
\end{align*}
where $S_{ij\varkappa}$ are real symmetric, and $M_{ij\varkappa}$ are real mixed-symmetry tensors. They depend on the indicated powers of $\bs{J}$. From the definitions given in \cite{itin2022decomposition}, we can derive the following tensors:

\begin{align}
\mathcal{S}^{\bs{J}}_{ij\varkappa}&=\tfrac{1}{5}\left(\mathcal{J}_i\delta_{j\varkappa}+\mathcal{J}_j\delta_{i\varkappa}+\mathcal{J}_\varkappa\delta_{ij}\right)\label{eq: rank3 tensors}\\
\mathcal{T}_{ij\varkappa}&=\tfrac{1}{6}(J_iJ_jJ_\varkappa +J_jJ_\varkappa J_i +J_\varkappa J_iJ_j+J_jJ_iJ_\varkappa+J_\varkappa J_jJ_i+J_iJ_\varkappa J_j)\nonumber\\
\mathcal{S}^{\bs{J}^3}_{ij\varkappa}&=\br{n}\Bigr[\mathcal{T}_{ij\varkappa}-\tfrac{1}{5}(\mathcal{T}_{inm}\delta_{nm}\delta_{j\varkappa}+\mathcal{T}_{njm}\delta_{nm}\delta_{i\varkappa}+\mathcal{T}_{nm\varkappa}\delta_{nm}\delta_{ij})\Bigl]\ke{n}\nonumber \\
\mathcal{M}^{\bs{J}}_{ij\varkappa}&=\tfrac{1}{4}(2\delta_{j\varkappa}\mathcal{J}_i-\delta_{ij}\mathcal{J}_\varkappa-\delta_{i\varkappa}\mathcal{J}_j)\nonumber\\
\mathcal{M}^{\bs{J}^2}_{ij\varkappa}&=\tfrac{1}{2}(\delta_{jl}\varepsilon_{im\varkappa}+\delta_{\varkappa l}\varepsilon_{imj})\mathcal{Q}_{lm}^{\bs{J}^2}\nonumber
\end{align}
So far, we have not used the fact that the quadrupole tensor is traceless. This reduces its number of independent elements by one, leading to $3\otimes 5=15$. For the tensor $\prescript{}{}{\alpha}_{ij\varkappa}^{\ell}$, this means that the partial trace over the last two components vanishes. $\mathcal{M}^{\bs{J}}_{ij\varkappa}$, $\mathcal{M}^{\bs{J}^2}_{ij\varkappa}$, and $\mathcal{S}^{\bs{J}^3}_{ij\varkappa}$ fullfill this condition. For the symmetric vector component $\mathcal{S}^{\bs{J}}_{ij\varkappa}$, we have however:
\begin{align*}
 0 = \prescript{}{}{\bar{\alpha}}_{v_s}^{\ell}\,\sum_l \mathcal{S}^{\bs{J}}_{ill}&=\prescript{}{}{\bar{\alpha}}_{v_s}^{\ell}\,\sum_l\tfrac{1}{5}\left(\mathcal{J}_i\delta_{ll}+\mathcal{J}_l\delta_{il}+\mathcal{J}_l\delta_{il}\right)=\prescript{}{}{\bar{\alpha}}_{v_s}^{\ell}\,\mathcal{J}_i
\end{align*}
This implies $\prescript{}{}{\bar{\alpha}}_{v_s}^{\ell}=0$. Therefore the contribution of $\mathcal{S}^{\bs{J}}_{ij\varkappa}$ vanishes and we arrive at the decomposition of the rank-3 tensor into only 3 irreducible structures:
\begin{align*}
\prescript{}{}{\alpha}_{ij\varkappa}^{\ell}=\mathcal{S}_{ij\varkappa}^{\bs{J}^3}\,\prescript{}{}{\bar{\alpha}}_{t_s}^{\ell}+\mathcal{M}_{ij\varkappa}^{\bs{J}}\,\prescript{}{}{\bar{\alpha}}_{v_m}^{\ell}+\mathcal{M}_{ij\varkappa}^{\bs{J}^2}\,\prescript{}{}{\bar{\alpha}}_{t_m}^{\ell}
\end{align*}
Again, the decomposition of ${\alpha'}_{ij\varkappa}^{\ell}$ follows completely analogous.

\subsubsection{Dipole-quadrupole terms: restrictions under parity and time reversal}\label{sec:electric dipole- electric quadrupole}
The restrictions from parity still apply as before: Only polarizabilities containing an even number of $P$-odd operators can exist.\\
To analyze the behavior under time reversal, we can proceed analogously to sec. \ref{sec:electric dipole- magnetic dipole}. There, we saw that if the polarizability contains an even number of $T$-odd operators, then the $T$-even scalar and tensor terms are real while the $T$-odd vector component is purely imaginary. As can be seen by the indicated powers in $J$, the rank-3 polarizability components proportional to $\mathcal{M}^{\bs{J}}_{ij\varkappa}$, and $\mathcal{S}^{\bs{J}^3}_{ij\varkappa}$ are $T$-odd, while the one proportional to $\mathcal{M}^{\bs{J}^2}_{ij\varkappa}$ is $T$-even. Therefore, following the same line of argument as before, we get:
\begin{align*}
\text{even number of $T$-odd operators: }&\prescript{}{}{\alpha}_{ij\varkappa}^{\ell}=i\mathcal{S}_{ij\varkappa}^{\bs{J}^3}\,\prescript{}{}{\alpha}_{t_s}^{\ell}+i\mathcal{M}_{ij\varkappa}^{\bs{J}}\,\prescript{}{}{\alpha}_{v_m}^{\ell}+\mathcal{M}_{ij\varkappa}^{\bs{J}^2}\,\prescript{}{}{\alpha}_{t_m}^{\ell}\\
\text{odd number of $T$-odd operators: }&\prescript{}{}{\alpha}_{ij\varkappa}^{\ell}=\mathcal{S}_{ij\varkappa}^{\bs{J}^3}\,\prescript{}{}{\alpha}_{t_s}^{\ell}+\mathcal{M}_{ij\varkappa}^{\bs{J}}\,\prescript{}{}{\alpha}_{v_m}^{\ell}+i\mathcal{M}_{ij\varkappa}^{\bs{J}^2}\,\prescript{}{}{\alpha}_{t_m}^{\ell}
\end{align*}
with $\prescript{}{}{\alpha}_{t_s}^{\ell},\prescript{}{}{\alpha}_{v_m}^{\ell},\prescript{}{}{\alpha}_{t_m}^{\ell} \in \mathbb{R}$.

\subsubsection{Rank-4 tensors}\label{sec: electric quadrupole- electric quadrupole}
The only rank-4 tensors that appear in our derivation are: $\prescript{QQ}{}{\alpha}_{ij\varkappa l}^{\ell}$, $\prescript{QQ}{P}{\alpha}_{ij\varkappa l}^{\ell}$, $\prescript{QQ}{PT}{\alpha}_{ij\varkappa l}^{\ell}$, and $\prescript{QQ}{T}{\alpha}_{ij\varkappa l}^{\ell}$. Without decomposing the tensor, we can show that $\prescript{QQ}{P}{\alpha}_{ij\varkappa l}^{\ell}$, $\prescript{QQ}{PT}{\alpha}_{ij\varkappa l}^{\ell}$, and $\prescript{QQ}{T}{\alpha}_{ij\varkappa l}^{\ell}$ vanish (see appendix \ref{sec:Insensitivity of dd and mumu terms to VN }). Because we are mainly interested in symmetry-violating polarizabilities, we will not enter deeper into the structure of $\prescript{QQ}{}{\alpha}_{ij\varkappa l}^{\ell}$. The decomposition of a tensor with the same symmetries as $\prescript{QQ}{}{\alpha}_{ij\varkappa l}^{\ell}$ can be found in \cite{itin2013constitutive}.
All of this applies to the \textit{primed} tensors $\prescript{QQ}{}{\alpha'}_{ij\varkappa l}^{\ell}$ in the same way.

\subsubsection{Insensitivity of $dd$, $\mu\mu$, and $QQ$ terms to $V^N$} \label{sec:Insensitivity of dd and mumu terms to VN }
From the arguments presented in sec. \ref{sec:electric dipole- magnetic dipole} one would expect $\prescript{dd}{T}{\alpha}_{ij}^{\ell}$, and $\prescript{dd}{T}{\alpha'}_{ij}^{\ell}$ to have nonvanishing components. The same goes for $\prescript{\mu\mu}{T}{\alpha}_{ij}^{\ell}$, $\prescript{\mu\mu}{T}{\alpha'}_{ij}^{\ell}$, $\prescript{QQ}{T}{\alpha}_{ij\varkappa l}^{\ell}$, and $\prescript{QQ}{T}{\alpha'}_{ij\varkappa l}^{\ell}$. In this section, we show that all polarizabilities quadratic in a multipole moment possess an additional symmetry that prevents the existence of $T$-odd terms. We demonstrate this on the example of $\prescript{dd}{T}{\alpha}_{ij}^{\ell}$:
\begin{align*}
  \prescript{d d}{T}{\alpha}_{ij}^{\ell}= 2 \sum_{k,\,l}\Bigg[ &\frac{\br{n}V^{T}\ke{l}\br{l}d_i\ke{k}\br{k}d_j\ke{n}}{(\omega_{kn}^2-(\omega^\ell)^2)\omega_{ln}}\omega_{kn}+\frac{\br{n}d_i\ke{k}\br{k}d_j\ke{l}\br{l}V^{T}\ke{n}}{(\omega_{kn}^2-(\omega^\ell)^2)\omega_{ln}}\omega_{kn}\nonumber\\
  +&\frac{\br{n}d_i\ke{k}\br{k}V^{T}\ke{l}\br{l}d_j\ke{n}}{(\omega_{kn}^2-(\omega^\ell)^2)(\omega_{ln}^2-(\omega^\ell)^2)}\left(\omega_{kn}\omega_{ln}+(\omega^\ell)^2\right)\Bigg]\nonumber\\
\end{align*}
From this definition, it follows that $\big(\prescript{d d}{T}{\alpha}_{ij}^{\ell}\big)^*=\prescript{d d}{T}{\alpha}_{ji}^{\ell}$. Regarding the tensor decomposition, this relation implies:
\begin{align}
 \left(\prescript{dd}{T}{\alpha}_{ij}^{\ell k}\right)^*=\delta_{ij}\; \prescript{dd}{T}{\bar{\alpha}}_s^{\ell k}-\varepsilon_{ij\varkappa}\mathcal{J}_\varkappa\; \prescript{dd}{T}{\bar{\alpha}}_v^{\ell k}+\mathcal{Q}^{\bs{J}^2}_{ij}\; \prescript{dd}{T}{\bar{\alpha}}_t^{\ell k}
\end{align}
where we used that $\varepsilon_{ij\varkappa}$ is antisymmetric.
However, from its transformation under time reversal, we know that $\prescript{dd}{T}{\alpha}_{ij}^{\ell}$, which contains an odd number of $T$-odd operators, must fulfill the relation:
\begin{align}
 \left(\prescript{dd}{T}{\alpha}_{ij}^{\ell k}\right)^*=-\delta_{ij}\; \prescript{dd}{T}{\bar{\alpha}}_s^{\ell k}+\varepsilon_{ij\varkappa}\mathcal{J}_\varkappa\; \prescript{dd}{T}{\bar{\alpha}}_v^{\ell k}-\mathcal{Q}^{\bs{J}^2}_{ij}\; \prescript{dd}{T}{\bar{\alpha}}_t^{\ell k}
\end{align}
Together, these relations can only be true if all three components are zero.\\
We can make analogous arguments for the polarizabilities that are quadratic in $\mu$.\\
In the case of $\prescript{QQ}{T}{\alpha}_{ij\varkappa l}^{\ell k}$, we are dealing with a rank-4 tensor. Because the quadrupole operator is symmetric, it obeys the relation $\left(\prescript{QQ}{}{\alpha}_{ij\varkappa l}^{\ell k}\right)^*=\prescript{QQ}{}{\alpha}_{\varkappa lij}^{\ell k}$. One can again see that this relation is in conflict with the relations that follow from time reversal. \\

In Ref. \cite{derevianko2010cp}, macroscopic equations were formulated describing the possible $P$, $PT$, and $T$ odd responses of systems to homogeneous time-dependent electromagnetic fields. From the results of this section, we can see that (under our initial assumptions) there are no microscopic processes that could generate the proposed $T$-odd, $P$-even responses.

\subsection{Derivation of $\langle V^P\rangle$, $\langle V^{PT}\rangle$, and $\langle V^T\rangle$} 
\label{app: Derivation of V^P V^PT V^T}

In the following, we illustrate how the values for the induced energy shifts $\langle V^P\rangle$, $\langle V^{PT}\rangle$, and $\langle V^T\rangle$ can be derived. \\
Obviously, all these potentials vanish in 0st order in $V^{EM}$:
\begin{align*}
 \br{n} V^P \ke{n}=\br{n} V^{PT} \ke{n}=\br{n} V^T \ke{n}=0
\end{align*}
If we now consider how $V^P$, $V^{PT}$, and $V^T$ are perturbed through $V^{EM}$, one could assume that the situation will be identical to above where we perturbed $V^{EM}$ through $V^P$, $V^{PT}$, and $V^T$. This is, however, only the case for purely static fields, as we can see in the following example:\\

We consider the perturbation of $V^P$ through $-\bs{d}\cdot\bs{\mathcal{E}}$. Such a system can be solved completely analogously to our derivation in sec. \ref{sec: perturbative treatment}, cf. Eq. (\ref{eq:NdtM2}), with the replacement $d_i\rightarrow V^P$:
\begin{align}
 \langle V^P \rangle \simeq
 2Re \sum_{k} 
 &\frac{\br{n}V^P\ke{k}\br{k}d_i\ke{n}}{\omega_{kn}^2-(\omega^\ell)^2}\omega_{kn}\mathcal{E}_{i}^\ell \label{eq: VP by d}\\
 -2Im&\frac{\br{n}V^P\ke{k}\br{k}d_i\ke{n}}{\omega_{kn}^2-(\omega^\ell)^2}\dot{\mathcal{E}}_{i}^\ell\nonumber
\end{align}
From our discussion in appendix \ref{sec: T-violation}, it follows that $\br{n}V^P\ke{k}\br{k}d_i\ke{n}$ must be fully imaginary.

While the real part (that is similar to the perturbation of $\bs{d}$ through $V^P$: $
 \langle \bs{d} \rangle\cdot \bs{\mathcal{E}}^\ell =
 2Re \sum_{k} 
 \br{n}d_i\ke{k}\br{k}V^P\ke{n} \omega_{kn}^{-1}\mathcal{E}_{i}^\ell $) vanishes, we still remain with a non-zero imaginary part:
\begin{align}
 \langle V^P \rangle \simeq
 -2Im \sum_{k} 
 &\frac{\br{n}V^P\ke{k}\br{k}d_i\ke{n}}{\omega_{kn}^2-(\omega^\ell)^2}\dot{\mathcal{E}}_{i}^\ell \equiv \prescript{d}{P}{\beta'}_{}^{\ell}\,\bs{\mathcal{J}}\cdot\dot{\bs{\mathcal{E}}}^\ell \label{eq: def beta}
\end{align}
From this, we can also conclude that even if the Lee-Yang argument (see appendix \ref{sec: T-violation}) forbids an atom to possess a static electric dipole moment without $T$-violation, this does not mean that there cannot be a purely $P$-odd linear response to an applied (time-dependent) electric field. \\

We find the following general rules: For $\beta$, $\beta'$ to be nonzero, they must contain an even number of $P$-odd operators. If they are rank-1 tensors, then an even number of $T$-odd operators makes them imaginary, while an odd number of $T$-odd operators makes them real. For rank-2 tensors, this is the other way around. These criteria lead to eq. (\ref{eq: <V>}) for the 1st order expressions for $\langle V^P\rangle$, $\langle V^{PT}\rangle$, and $\langle V^T\rangle$.\\

Deriving the 2nd order perturbations through $V^{EM}$ is more complicated. This is because we have two time-dependent perturbers, which, through interferences, create a multitude of terms (see, for example, Refs. \cite{langhoff1972aspects, boyd2008nonlinear,mandal2012adiabatic}). For this reason, we limit our discussion of the 2nd order corrections of $\langle V^P\rangle$, $\langle V^{PT}\rangle$, and $\langle V^T\rangle$ to the time-independent case. In appendix \ref{sec: 2nd order in VEM}, we demonstrate that the resulting terms coincide with those in sec. \ref{sec: Full expressions of the induced moments}.
\section{Derivation of the $P$, $T$ violating polarizability expressions }
\label{sec: Simultaneous perturbation through electromagnetic an symmetry violating potentials}
In the following, we derive the expressions for the polarizabilities that originate from the simultaneous perturbation through the electromagnetic potential $V^{EM}$ and the symmetry-violating potential $V^N$.

\subsection{0st and 1st order in $V^{EM}$}\label{sec: 1th order in VEM}
Because $V^{EM}$ is time dependent, we need to solve the system using time-dependent perturbation theory. We can, for example, define the perturbation potential $V=-\mu_i \mathcal{H}_i-V^{PT}$ and solve the 2nd-order correction according to Ref. \cite{mandal2012adiabatic}. We ignore here and throughout the paper terms that are quadratic in $V^N$ due to their small size. If we, for now, also ignore terms that are quadratic in $V^{EM}$, we obtain the following expression for the perturbed time-dependent wavefunction:
\begin{align*}
  \ke{N(t)} \simeq \ke{n}+\sum_k &\frac{\br{k}\mu_i\ke{n}}{\omega_{kn}^2-(\omega^\ell)^2}\left[\omega_{kn}\mathcal{H}_{i}^\ell+i\dot{\mathcal{H}}_{i}^\ell\right]\ke{k}+\sum_k \frac{\br{k}V^{PT}\ke{n}}{\omega_{kn}}\ke{k}\\
  +\sum_{k,\,l}&\frac{\br{k}\mu_i\ke{l}\br{l}V^{PT}\ke{n}}{(\omega_{kn}^2-(\omega^\ell)^2)\omega_{ln}}\left[\omega_{kn}\mathcal{H}_{i}^\ell+i\dot{\mathcal{H}}_{i}^\ell\right]\ke{k}\\
  +\sum_{k,\,l}&\frac{\br{k}V^{PT}\ke{l}\br{l}\mu_i\ke{n}}{(\omega_{kn}^2-(\omega^\ell)^2)\omega_{ln}}\ke{k}\Bigg]\left[\left(\omega_{kn}\omega_{ln}+(\omega^\ell)^2\right)\mathcal{H}_{i}^\ell+\left(\omega_{kn}+\omega_{ln}\right)\dot{\mathcal{H}}_{i}^\ell\right]\\
  -\sum_k& \frac{\br{k}V^{PT}\ke{n}\br{n}\mu_i\ke{n}}{\omega_{kn}^2}\ke{k}\,\mathcal{H}_{i}^\ell
\end{align*}
Here we already removed a term that is proportional to $\br{n}V^{PT}\ke{n}=0$. \\
From this, we can now calculate the 1st order (in $V^{EM}$) perturbation of the $P$, $T$ odd induced electric dipole moment:

\begin{align}
\langle d_i \rangle =\br{N(t)}&d_i\ke{N(t)} \simeq 2Re\sum_{k}\Bigg[ \frac{\br{n}V^{PT}\ke{k}\br{k}d_i\ke{n}}{\omega_{kn}}\Bigg]\label{eq: alpha d mu}\\
  + 2Re\sum_{k,\,l}\Bigg[ &\frac{\br{n}V^{PT}\ke{l}\br{l}d_i\ke{k}\br{k}\mu_j\ke{n}}{(\omega_{kn}^2-(\omega^\ell)^2)\omega_{ln}}\omega_{kn}+\frac{\br{n}d_i\ke{k}\br{k}\mu_j\ke{l}\br{l}V^{PT}\ke{n}}{(\omega_{kn}^2-(\omega^\ell)^2)\omega_{ln}}\omega_{kn} \nonumber\\
  +&\frac{\br{n}d_i\ke{k}\br{k}V^{PT}\ke{l}\br{l}\mu_j\ke{n}}{(\omega_{kn}^2-(\omega^\ell)^2)(\omega_{ln}^2-(\omega^\ell)^2)}\left(\omega_{kn}\omega_{ln}+(\omega^\ell)^2\right)\Bigg]\mathcal{H}_{j}^\ell\nonumber\\
  - 2Im\sum_{k,\,l}\Bigg[ &\frac{\br{n}V^{PT}\ke{l}\br{l}d_i\ke{k}\br{k}\mu_j\ke{n}}{(\omega_{kn}^2-(\omega^\ell)^2)\omega_{ln}}+\frac{\br{n}d_i\ke{k}\br{k}\mu_j\ke{l}\br{l}V^{PT}\ke{n}}{(\omega_{kn}^2-(\omega^\ell)^2)\omega_{ln}}\nonumber\\
  +&\frac{\br{n}d_i\ke{k}\br{k}V^{PT}\ke{l}\br{l}\mu_j\ke{n}}{(\omega_{kn}^2-(\omega^\ell)^2)(\omega_{ln}^2-(\omega^\ell)^2)}\left(\omega_{ln}+\omega_{kn}\right)\Bigg]\dot{\mathcal{H}}_{j}^\ell\nonumber\\
  -2Re\sum_k&\frac{\br{n}d_i\ke{k}\br{k}V^{PT}\ke{n}}{\omega_{kn}^2}\br{n}\mu_i\ke{n}\,\mathcal{H}_{j}^\ell\nonumber\\
  \equiv&Re\Big[ \prescript{d}{PT}{\alpha}_{i}\Big]+Re\Big[\prescript{d\mu}{PT}{\alpha}_{ij}^{\ell }\Big]\mathcal{H}_{j}^\ell+Im\Big[\prescript{d\mu}{PT}{\alpha'}_{ij}^{\ell }\Big]\dot{\mathcal{H}}_{j}^\ell+R\nonumber
\end{align}
Here, we ignored the term $\prescript{d\mu}{}{\alpha}_{ij}^{\ell }$, which vanishes due to parity.\\
We can see that the solution to this problem not only gives us expressions for the $P$, $T$ odd polarizabilities but also leads to a certain type of rest term:
\begin{align*}
R=-2Re\sum_k&\frac{\br{n}d_i\ke{k}\br{k}V^{PT}\ke{n}}{\omega_{kn}^2}\br{n}\mu_j\ke{n}\,\mathcal{H}_{j}^\ell
\end{align*}
Depending on the time-dependent perturbation treatment, these kinds of rest terms can be absent (see, for example, Ref. \cite{boyd2008nonlinear}). They are, however, necessary for a proper normalization and convergence in the limit of static fields \cite{langhoff1972aspects, mandal2012adiabatic}.\\

It is apparent that in the rest term, $\bs{d}$ and $\bs{\mu}$ are uncorrelated in the sense that their expectation values factorize. For example, $\mu_j$ in the expression above does not impose any restrictions on the $\ke{k}$-states that $d_i$ and $V^{PT}$ connect to. For these reasons, it makes sense to view these as products of lower-order terms. As such, they only provide small corrections to already considered observables without introducing the new tensor structures that we saw in the $\prescript{}{}{\alpha}_{ij}^{\ell}$. For these reasons, we ignore these $R$-terms elsewhere.
\subsection{2nd order in $V^{EM}$} \label{sec: 2nd order in VEM} 
In this order of perturbation, we will only consider the expectation values of the $P$, $T$ odd Potentials $V^P$, $V^{PT}$, and $V^T$. These terms that are linear in $V^N$ but quadratic in $V^{EM}$ will, therefore, be of the same size as the ones we discussed above. As mentioned before, a time-dependent treatment in this order is more involved and beyond the scope of this paper. We will only discuss the time-independent case. We get, for example, for the perturbation of $\langle V^P \rangle$ through $d$, and $\mu$:
\begin{align}
 2Re \sum_{k,\,l}
 \Bigg[&\frac{\br{n}d_i\ke{l}\br{l}V^P\ke{k}\br{k}\mu_j\ke{n}}{\omega_{kn}\omega_{ln}}+\frac{\br{n}V^{PT}\ke{l}\br{l}d_i\ke{k}\br{k}\mu_j\ke{n}}{\omega_{kn}\omega_{ln}}\\+&\frac{\br{n}V^{PT}\ke{k}\br{k}\mu_j\ke{l}\br{l}d_i\ke{n}}{\omega_{kn}\omega_{ln}}\Bigg]\mathcal{E}_{i}\mathcal{H}_{j}\nonumber\\
 &\equiv Re\Big[\prescript{d\mu}{P}{\beta}_{ij}^{}\Big]\mathcal{E}_{i}\mathcal{H}_{j}\nonumber
\end{align}
This is identical to (\ref{eq: alpha PT d mu}) in the limit $\ell= 0$. Therefore it follows $\prescript{d\mu}{P}{\beta}_{ij}^{}=\prescript{d\mu}{P}{\alpha}_{ij}^{0}\equiv\prescript{d\mu}{P}{\alpha}_{ij}^{}$.

\section{Energy shift from static electromagnetic fields} \label{sec: Energy shift from static electromagnetic fields}
We now derive how the polarizations are related to the energy shift that is induced when static electromagnetic fields are applied to the system. As before, our system is described by $H=H^0+V^{EM}+V^N$. The perturbative energy corrections in such a situation are well known \cite{landau2013quantum6,mandal2012adiabatic} and given by:
\begin{align*}
\Delta E\; &\simeq \Delta E_1+\Delta E_2+\Delta E_3\\
  \Delta E_1&=\br{n}\left( V^{EM}+V^N\right)\ke{n}\\
  \Delta E_2&=-\sum_k \frac{|\br{n}\left( V^{EM}+V^N\right)\ke{k}|^2}{\omega_{kn}}\\
   \Delta E_3&=\sum_{k,\,l} \frac{\br{n}\left( V^{EM}+V^N\right)\ke{k}\br{k}\left( V^{EM}+V^N\right)\ke{l}\br{l}\left( V^{EM}+V^N\right)\ke{n}}{\omega_{kn}\,\omega_{ln}}\\
   &\quad -\br{n}\left( V^{EM}+V^N\right)\ke{n}\sum_k \frac{|\br{n}\left( V^{EM}+V^N\right)\ke{k}|^2}{\omega_{kn}^2}
\end{align*}
As discussed in appendix \ref{sec: 0st order}, most terms in $\Delta E_1$ vanish:
\begin{align*}
  \Delta E_1= -\br{n}\mu_i\ke{n}\mathcal{H}_i-\br{n}Q_{ij}\ke{n}\partial_i\mathcal{E}_j=-\prescript{\mu}{}{\alpha}\,\mathcal{J}_i\mathcal{H}_i-\prescript{Q}{}{\alpha}\,\mathcal{Q}^{\bs{J}^2}_{ij}\partial_i\mathcal{E}_j
\end{align*}
In the next order, we get both the $P$, $T$ odd polarizabilities that are linear in $V^{EM}$ (see sec. \ref{sec: Induced V^P V^PT V^T}) and the $P$, $T$ even polarizabilities that are quadratic in $V^{EM}$ (see sec. \ref{sec: Time dependent perturbation}). For example for $V^{EM'}=-\bs{d}\cdot \bs{\mathcal{E}}$, and $V^{N'}=-V^{PT}$, we get:
\begin{align*}
  \Delta E_2&=-\sum_k \frac{\br{n}d_i\ke{k} \br{k}d_j\ke{n}}{\omega_{kn}}\mathcal{E}_i\mathcal{E}_j- \frac{\br{n}d_i\ke{k} \br{k}V^{PT}\ke{n}}{\omega_{kn}}\mathcal{E}_i- \frac{\br{n}V^{PT}\ke{k} \br{k}d_i\ke{n}}{\omega_{kn}}\mathcal{E}_i\\
  &=-\tfrac{1}{2}\prescript{dd}{}{\alpha}_{s}\, \mathcal{E}^2-\tfrac{1}{2}\prescript{dd}{}{\alpha}_{t}\, \mathcal{Q}^{\bs{J}^2}_{ij}\,\mathcal{E}_i\mathcal{E}_j-\prescript{d}{PT}{\alpha}_{}^{}\, \mathcal{J}_i \mathcal{E}_i
\end{align*}
The 3rd energy correction $\Delta E_3$ is a sum of two terms. The second one describes the shift from the rest terms that we discussed in appendix \ref{sec: 1th order in VEM}. For the reasons stated before, we neglect these. The first term in $\Delta E_3$ gives us the $P$, $T$-odd polarizabilities quadratic in $V^{EM}$. From the example $V^{EM'}=-\bs{d}\cdot \bs{\mathcal{E}}-\bs{\mu}\cdot \bs{\mathcal{H}}$, and $V^{N'}=-V^{PT}$, we get:
\begin{align*}
   \Delta E_3=-\sum_{k,\,l} [&(\omega_{kn}\omega_{ln})^{-1}\, \mathcal{E}_i\mathcal{H}_j]\\
   \cdot\Big[&\br{n}d_i\ke{k}\br{k}\mu_j\ke{l}\br{l}V^{PT}\ke{n}
  + \br{n}d_i\ke{k}\br{k}V^{PT}\ke{l}\br{l}\mu_j\ke{n}\\
   +&\br{n}\mu_j\ke{k}\br{k}d_i\ke{l}\br{l}V^{PT}\ke{n}
   +\br{n}\mu_j\ke{k}\br{k}V^{PT}\ke{l}\br{l}d_i\ke{n}\\
   +&\br{n}V^{PT}\ke{k}\br{k}d_i\ke{l}\br{l}\mu_j\ke{n}
     +\br{n}V^{PT}\ke{k}\br{k}\mu_j\ke{l}\br{l}d_i\ke{n}\Big]\\
     =-\prescript{d\mu}{PT}{\alpha}_{s}\, \mathcal{E}_i&\mathcal{H}_i-\prescript{d\mu}{PT}{\alpha}_{t}\, \mathcal{Q}^{\bs{J}^2}_{ij}\,\mathcal{E}_i\mathcal{H}_j
\end{align*}
From this, we can also see: $\prescript{d\mu}{PT}{\alpha}_{s} = \prescript{\mu d}{PT}{\alpha}_{s}$.
If we perform the treatment for the full $V^N$ and $V^{EM}$, we arrive at a general expression for the energy shift of an atom or molecule exposed to static electromagnetic fields:
\begin{align}
 \Delta E\simeq& -\prescript{\mu}{}{\alpha}_{}\, \bs{\mathcal{J}}\cdot\bs{\mathcal{\mathcal{H}}} -\tfrac{1}{2} \prescript{dd}{}{\alpha}_{s}\, {\mathcal{E}}^2 -\tfrac{1}{2} \prescript{\mu\mu}{}{\alpha}_{s}\, {\mathcal{H}}^2 - \tfrac{1}{2}\prescript{dd}{}{\alpha}_{t}\, \mathcal{Q}^{\bs{J}^2}_{ij} {\mathcal{E}}_{i} {\mathcal{E}}_{j} - \tfrac{1}{2}\prescript{\mu \mu}{}{\alpha}_{t}\, \mathcal{Q}^{\bs{J}^2}_{ij} {\mathcal{H}}_{i} {\mathcal{H}}_{j}\nonumber\\
 & -\prescript{d}{PT}{\alpha}_{}^{}\, \bs{\mathcal{J}}\cdot\bs{\mathcal{E}} - \prescript{d\mu}{PT}{\alpha}_{s}^{}\, \bs{\mathcal{H}}\cdot\bs{\mathcal{E}} -\prescript{d\mu}{P}{\alpha}_{v}^{}\, (\bs{\mathcal{E}}\times\bs{\mathcal{H}})\cdot \bs{\mathcal{J}}- \prescript{d\mu}{PT}{\alpha}_{t}^{}\, \mathcal{Q}^{\bs{J}^2}_{ij}\mathcal{E}_i{\mathcal{H}}_j\nonumber\\
 &-\prescript{Q}{}{\alpha}_{}^{}\,\mathcal{Q}^{\bs{J}^2}_{ij}\partial_i \mathcal{E}_j -\tfrac{1}{2}\prescript{QQ}{}{\alpha}_{ij\varkappa l}\,\partial_i \mathcal{E}_j\partial_\varkappa \mathcal{E}_l\nonumber\\
 &-\left(\prescript{\mu Q}{}{\alpha}_{t_s}\,\mathcal{S}_{ij\varkappa}^{\bs{J}^3}+\prescript{\mu Q}{}{\alpha}_{v_m}\,\mathcal{M}_{ij\varkappa}^{\bs{J}}\right)\mathcal{H}_i \partial_j {\mathcal{E}}_\varkappa\nonumber\\
 & -\prescript{dQ}{P}{\alpha}_{t_m}^{}\,\mathcal{M}_{ij\varkappa}^{\bs{J}^2}\mathcal{E}^{}_i \partial_j {\mathcal{E}}_\varkappa-\prescript{\mu Q}{T}{\alpha}_{t_m}^{}\,\mathcal{M}_{ij\varkappa}^{\bs{J}^2}\mathcal{H}^{}_i \partial_j {\mathcal{E}}_\varkappa\nonumber\\
 &-\left(\prescript{d Q}{PT}{\alpha}_{t_s}^{\ell}\,\mathcal{S}_{ij\varkappa}^{\bs{J}^3}+\prescript{d Q}{PT}{\alpha}_{v_m}^{}\,\mathcal{M}_{ij\varkappa}^{\bs{J}}\right)\mathcal{E}^{}_i \partial_j {\mathcal{E}}_\varkappa\nonumber
\end{align}
\section{Alignment of $\mathcal{M}_{ij\varkappa}^{\bs{J}}$, and $\mathcal{M}_{ij\varkappa}^{\bs{J}^2}$ in externally applied fields}\label{sec:Alignment of MM in externally applied fields}

In sec. \ref{sec: Static inhomogeneous fields}, we discuss observables in static inhomogeneous fields. They rely on the angular momentum tensor structures $\mathcal{M}_{ij\varkappa}^{\bs{J}}$, and $\mathcal{M}_{ij\varkappa}^{\bs{J}^2}$. In this section, we show how these can get oriented under the application of externally applied fields. \\

The tensor $\mathcal{M}_{ij\varkappa}^{\bs{J}^2}$ depends on $\mathcal{Q}^{\bs{J}^2}_{ij}$ (defined in (\ref{eq: def Q})) in the following way:
\begin{align}
\mathcal{M}^{\bs{J}^2}_{ij\varkappa}&=\tfrac{1}{2}(\delta_{jl}\varepsilon_{im\varkappa}+\delta_{\varkappa l}\varepsilon_{imj})\mathcal{Q}_{lm}^{\bs{J}^2}\label{eq: MJ2}
\end{align}
Similar to the alignment of $\bs{\mathcal{J}}$ into the direction of an applied magnetic field discussed in sec. \ref{sec: Alignment of J in externally applied fields}, $\mathcal{Q}^{\bs{J}^2}_{ij}$ as well can be aligned through  $P$-even $T$-even interactions and sufficient field strengths. In this case, all fields $\bs{\mathcal{E}}$, $\bs{\mathcal{H}}$, and $\partial_i\mathcal{E}_j$ compete with each other to define the specific state $\ke{aJm}$ that minimizes the energy and indirectly defines the orientation of $\mathcal{Q}^{\bs{J}^2}_{ij}$. We can express this by performing the replacement
\begin{align}
 \mathcal{Q}^{\bs{J}^2}_{ij}\rightarrow J^2\frac{b_1\partial_i \mathcal{E}_j+b_2 {\mathcal{E}}_{i} {\mathcal{E}}_{j} +b_3{\mathcal{H}}_{i} {\mathcal{H}}_{j}}{\sqrt{\sum_{\varkappa,l} \left(b_1\partial_\varkappa \mathcal{E}_l+ b_2{\mathcal{E}}_{\varkappa} {\mathcal{E}}_{l} +b_3{\mathcal{H}}_{\varkappa} {\mathcal{H}}_{l}\right)^2}}-\tau\delta_{ij}\label{eq: quadrupole allignement}
\end{align}
Here $b_1$, $b_2$, and $b_3$ are real constants that depend on the sizes of the different classically allowed polarizabilities. The expression has been normalized through the Frobenius norm. $\tau$ represents the trace of the first term. We need to subtract it, to ensure that the new object still remains traceless. \\

We now proceed by deriving the energy shifts and induced dipole moments that originate from terms proportional to $\mathcal{M}_{ij\varkappa}^{\bs{J}^2} \partial_j \mathcal{E}_\varkappa$.\\
The quadrupole component of an electrostatic field is given by the potential: $\phi^Q=q[(\bs{r}\cdot\hat{\bs{n}})^2-\tfrac{1}{3}\bs{r}^2]$. It is oriented along some axis $\hat{\bs{n}}$ and has an amplitude of $q$. without loss of generality, we can take $\hat{\bs{n}}$ in $z$-direction: $\hat{\bs{n}}=\hat{e}_z$. It follows for the electric field Jacobian:
\begin{align*}
\partial_i\mathcal{E}_j|_{\bs{r}=0}=-\partial_i\partial_j \phi|_{\bs{r}=0}=-\partial_i\partial_j \phi^Q=-2q[\delta_{iz}\delta_{jz}-\tfrac{1}{3}\delta_{ij}]
\end{align*}
From Eqs. (\ref{eq: MJ2}), (\ref{eq: quadrupole allignement}), and (\ref{eq: Delta E inhomo}) we can now derive the energy shift that is induced by the $\prescript{dQ}{P}{\alpha}_{t_m}^{}$ term: 
\begin{align*}
 \Delta E&=\,\prescript{dQ}{P}{\alpha}_{t_m}^{}\,q\,(\delta_{jl}\varepsilon_{im\varkappa}+\delta_{\varkappa l}\varepsilon_{imj})(\delta_{jz}\delta_{\varkappa z}-\tfrac{1}{3}\delta_{j\varkappa})\mathcal{Q}_{lm}\mathcal{E}_i\nonumber\\
 &= 2\,\prescript{dQ}{P}{\alpha}_{t_m}^{}\,q\,\varepsilon_{imz}\mathcal{Q}_{zm}\mathcal{E}_i\nonumber \\
 &=2 \,\prescript{dQ}{P}{\alpha}_{t_m}^{}\,q \mathcal{J}^2\,\varepsilon_{imz}\Bigg[J^2\frac{b_1\partial_i \mathcal{E}_j+b_2 {\mathcal{E}}_{i} {\mathcal{E}}_{j} +b_3{\mathcal{H}}_{i} {\mathcal{H}}_{j}}{\sqrt{\sum_{\varkappa,l} \left(b_1\partial_\varkappa \partial_l\phi^Q+ b_2{\mathcal{E}}_{\varkappa} {\mathcal{E}}_{l} +b_3{\mathcal{H}}_{\varkappa} {\mathcal{H}}_{l}\right)^2}}-\tau\delta_{ij}\Bigg]\mathcal{E}_i\nonumber
\end{align*}
From the antisymmetry of $\varepsilon_{ij\varkappa}$, it follows that $\varepsilon_{imz}\mathcal{E}_i\partial_z\partial_m \phi^Q$, $\varepsilon_{imz} {\mathcal{E}}_{z} {\mathcal{E}}_{m}\mathcal{E}_i$, and $\varepsilon_{ij\varkappa}\delta_{ij}\tau$ necessarily vanish. We can write the resulting energy shift as:
\begin{align*}
 \Delta E=2J^2 \frac{\prescript{dQ}{P}{\alpha}_{t_m}^{}\,b_3\, q\,(\bs{\mathcal{H}}\cdot \hat{\bs{n}})(\bs{\mathcal{E}}\times\bs{\mathcal{H}})\cdot \hat{\bs{n}}}{\sqrt{\sum_{\varkappa,l} \left(b_1\partial_\varkappa \partial_l\phi^Q+ b_2{\mathcal{E}}_{\varkappa} {\mathcal{E}}_{l} +b_3{\mathcal{H}}_{\varkappa} {\mathcal{H}}_{l}\right)^2}}+\dotsb
\end{align*}
The same interaction also leads to an induced electric dipole moment:
\begin{align*}
 \langle \bs{d} \rangle = -2J^2 \frac{\,\prescript{dQ}{P}{\alpha}_{t_m}^{}\,b_3\, q\,(\bs{\mathcal{H}}\cdot \hat{\bs{n}})(\bs{\mathcal{H}}\times \hat{\bs{n}})}{\sqrt{\sum_{\varkappa,l} \left(b_1\partial_\varkappa \partial_l\phi^Q+ b_2{\mathcal{E}}_{\varkappa} {\mathcal{E}}_{l} +b_3{\mathcal{H}}_{\varkappa} {\mathcal{H}}_{l}\right)^2}}+\dotsb
\end{align*}
The second $ \mathcal{M}^{\bs{J}^2}_{ij\varkappa}$-dependent term in Eq. (\ref{eq: Delta E inhomo}) has a similar structure. For it, we find:
\begin{align*}
 \Delta E &= 2J^2\frac{\prescript{\mu Q}{T}{\alpha}_{t_m}^{}\,b_2\,q\,(\bs{\mathcal{E}}\cdot \hat{\bs{n}})(\bs{\mathcal{H}}\times\bs{\mathcal{E}})\cdot \hat{\bs{n}}}{\sqrt{\sum_{\varkappa,l} \left(b_1\partial_\varkappa \partial_l\phi^Q+ b_2{\mathcal{E}}_{\varkappa} {\mathcal{E}}_{l} +b_3{\mathcal{H}}_{\varkappa} {\mathcal{H}}_{l}\right)^2}}+\dotsb\\
 \text{and }\langle \bs{\mu} \rangle &=-2J^2\frac{\prescript{ \mu Q}{T}{\alpha}_{t_m}^{}\,b_2\,q\,(\bs{\mathcal{E}}\cdot \hat{\bs{n}})(\bs{\mathcal{E}}\times \hat{\bs{n}})}{\sqrt{\sum_{\varkappa,l} \left(b_1\partial_\varkappa \partial_l\phi^Q+ b_2{\mathcal{E}}_{\varkappa} {\mathcal{E}}_{l} +b_3{\mathcal{H}}_{\varkappa} {\mathcal{H}}_{l}\right)^2}} +\dotsb\nonumber
\end{align*}

The energy shift $\Delta E$ in Eq. (\ref{eq: Delta E inhomo}) further contains two $PT$-odd terms. The first of them, couples the electric field $\mathcal{E}_i$ and the Hessian $\partial_j\partial_\varkappa \phi^Q$ to the octupole angular momentum structure $\mathcal{S}_{ij\varkappa}^{\bs{J}^3}$. Similar to the quadrupole in Eq. (\ref{eq: quadrupole allignement}), this octupole as well can be aligned through the application of external fields. 

The other $PT$-odd term in Eq. (\ref{eq: Delta E inhomo}) depends on the symmetric vector quantity:
\begin{align*}
 \mathcal{M}^{\bs{J}}_{ij\varkappa}&=\tfrac{1}{4}(2\delta_{j\varkappa}\mathcal{J}_i-\delta_{ij}\mathcal{J}_\varkappa-\delta_{i\varkappa}\mathcal{J}_j)
\end{align*}
We can derive a simplified form for $\mathcal{M}^{\bs{J}}_{ij\varkappa}$ similar to what we did for $\mathcal{M}^{\bs{J}^2}_{ij\varkappa}$:
\begin{align*}
 \Delta E &= q\,\prescript{d Q}{PT}{\alpha}_{v_m}^{}\,\tfrac{1}{2}(2\delta_{j\varkappa}\mathcal{J}_i-\delta_{ij}\mathcal{J}_\varkappa-\delta_{i\varkappa}\mathcal{J}_j) (\delta_{iz}\delta_{jz}-\tfrac{1}{3}\delta_{ij})\mathcal{E}_i \\
 &=q\,\prescript{d Q}{PT}{\alpha}_{v_m}^{}\,\left[(\bs{\mathcal{E}}\cdot \bs{\mathcal{J}})- (\bs{\mathcal{E}}\cdot \hat{\bs{n}})(\bs{\mathcal{J}}\cdot \hat{\bs{n}})\right]
\end{align*}
As discussed in sec. \ref{sec: Alignment of J in externally applied fields}, $\bs{\mathcal{J}}$ can align to the direction of an applied magnetic field, leading to:
\begin{align*}
 \Delta E &=q\,\prescript{d Q}{PT}{\alpha}_{v_m}^{}\,J\left[(\bs{\mathcal{E}}\cdot \bs{\mathcal{H}})- (\bs{\mathcal{E}}\cdot \hat{\bs{n}})(\bs{\mathcal{H}}\cdot \hat{\bs{n}})\right]+\dotsb\\
 \langle \bs{d} \rangle &= -q\,\prescript{d Q}{PT}{\alpha}_{v_m}^{}\, J\left[\bs{\mathcal{H}}- \hat{\bs{n}}(\bs{\mathcal{H}}\cdot \hat{\bs{n}})\right]+\dotsb
\end{align*}
\footnotesize
\bibliographystyle{ieeetr}
\bibliography{References}

\end{document}